\documentclass{jnmp01b}

\usepackage{longtable}

\numberwithin{equation}{section}

\def\be{\begin{equation}}
\def\ee{\end{equation}}
\def\bq{\begin{eqnarray}}
\def\eq{\end{eqnarray}}
\def\beq{\begin{eqnarray*}}
\def\eeq{\end{eqnarray*}}

\def\ds{\displaystyle}

\def\a{\alpha}

\def\la{\lambda}

\makeatletter
\def\LT@makecaption#1#2#3{%
  \LT@mcol\LT@cols c{\hbox to\z@{\hss\parbox[t]\LTcapwidth{%
    \sbox\@tempboxa{#1{\small{\textbf{#2}: }#3}}%
    \ifdim\wd\@tempboxa>\hsize
      #1{\small{\textbf{#2}: }#3}%
    \else
      \hbox to\hsize{\hfil\box\@tempboxa\hfil}%
    \fi
    \endgraf\vskip\baselineskip}%
  \hss}}}
\makeatother

\begin{document}

\renewcommand{\evenhead}{L.\ Cair\'o}
\renewcommand{\oddhead}{Integrability of the 3D Lotka-Volterra System}


\thispagestyle{empty}

\begin{flushleft}
\footnotesize \sf
Journal of Nonlinear Mathematical Physics \qquad 2000, V.7, N~4,
\pageref{firstpage}--\pageref{lastpage}.
\hfill {\sc Article}
\end{flushleft}

\vspace{-5mm}

\copyrightnote{2000}{L.\ Cair\'o}

\Name{Darboux First Integral Conditions and Integrability of the 3D
Lotka-Volterra System}

\label{firstpage}

\Author{Laurent CAIR\'O}

\Adress{MAPMO--UMR 6628.
Universit\'e d'Orl\'eans- UFR Sciences BP 6759\\
45067 Orl\'eans, C\'edex 2, France \\
email: lcairo@labomath.univ-orleans.fr}

\Date{Received March 14, 2000; Accepted May 5, 2000}

\begin{abstract}
\noindent
We apply the Darboux theory of integrability to polynomial ODE's of dimension $3$. 
Using this theory and computer algebra, we study the existence of first integrals 
for the $3$--dimensional Lotka--Volterra systems with polynomial invariant algebraic 
solutions linear and quadratic and determine numerous cases of integrability.
\end{abstract}


\section{Introduction}

In this paper we are concerned with the {\it $3$-dimensional Lotka-Volterra [1,2] 
system (LV3)}, having first integrals formed with linear and quadratic polynomials.
 The LV3 is
\be\label{l1}
\begin{array}{l}
{\dot{x}}=x(a_1 + b_{11} x + b_{12} y + b_{13} z) = P_1(x,y,z),\\[1ex]
{\dot{y}}=y(a_2 + b_{21} x + b_{22} y + b_{23} z) = P_2(x,y,z),\\[1ex]
{\dot{z}}=z(a_3 + b_{31} x + b_{32} y + b_{33} z) = P_3(x,y,z),
\end{array}
\ee
where the overdot means time derivative. 
This system typically model the time evolution of conflicting species in chemistry and
biology [3], where the linear terms $a_i$ (also called Malthusian terms)
denote growth (or decay) rates of each species independently of the others; the 
self-interactive terms, also called Verhulst terms ($b_{ii}$) 
represent the control on over-population of each of the respective species (such 
as cannibalism or depletion of resources), and the cross-interactive terms ($b_{ij}, i\neq j$)
represent inter-species interactions (such as predator-prey). It has been
extensively studied, starting with the pioneer works of Lotka [1] and Volterra [2], in the 
case where all the Verhulst terms vanish. Then in three dimensions the essential 
parameters reduce to only 6. A detailed study of this case can be found in Ref. 4
   
There are many other natural phenomena modeled by (\ref{l1}), such as the coupling 
of waves in laser physics [5], the evolution of electrons, ions and neutral 
species in plasma physics [6]. 
In hydrodynamics, they model the convective instability in the B\'enard
problem [7]. Similarly, they appear in the interaction of gases in a 
background host medium [8]. In the theory of partial differential 
equations they can be obtained as a discretized form of the Korteweg-de Vries 
equation [9]. They also play a role in such diverse topics of current 
interest as neural networks [10], biochemical reactions etc. The systems interest
became crucial after the work of Brenig and Goriely [11,12] wherein they
prove that a large class of ordinary differential equations implied in various 
fields of physics, biology, chemistry and economics, can be transformed into a 
Lotka-Volterra of greater order using a quasimonomial formalism. In the context of
plasma physics, all the nonlinear terms represent binary interactions or model 
certain transport across the boundary of the system.

In solving for the LV3, it is worthwhile to know, 
given a set of initial conditions, what its long-time asymptotic behaviour
 will be or whether stable periodic solutions exist. The existence of stable 
periodic orbits would be rather important for experimentalists wishing to obtain 
and maintain a stable oscillatory state. Since in general the solutions of the 
LV3 cannot be written in terms of elementary functions, the two questions of
 asymptotic behaviour and the existence of periodic orbits are rather hard to 
answer. Nor it is easy to explore the general solution using numerical schemes 
since one has to prescribe all the parameters of the LV3 in terms of real numbers.
 It will then be of most interest to possess {\it constants of the motion},
 which, as such, contain the trajectories, and permits the localisation of
 the solutions
or find the asymptotic state.  Particularly, if a three-dimensional dynamical system
admits a constant of motion, then the phase space is foliated into two-dimensional
 leaves and therefore certain types of irregular orbits cannot occur.
In some cases it is even possible to find for a given three-dimensional dynamical
 system two functionally independent constants of the motion. Then the orbits are 
the intersections of two-dimensional flow invariant solutions and therefore
nonchaotic. (Indeed, chaotic behaviour is associated with nonintegrability
of the dynamical system.) In fact, obtaining a constant of motion corresponds to a partial
 integration and is interesting both from an analytical and numerical point of 
view. In this last point of view, obtaining a constant of motion is equivalent to 
reducing by one unit the dimension of the phase space. Since in nonlinear 
problems we are usually interested in a full exploration of initial conditions,
 any reduction of the dimension corresponds to a dramatic saving of numerical
 computation. Moreover, the existence of a constant of motion provides a welcome check
 of the numerical scheme with respect to its accuracy and stability. 
 
To introduce the adopted terminology, we must note that a constant of motion 
may be time-dependent. Usually this
time-dependency appears in an exponential form [13,14]. Some authors [14] 
use the name of {\it invariants}, name which they apply regardless
of the type of constant of motion. As here we are concerned only with
time-independent constants of motion, we will keep for them the more usual name
of {\it first integrals}.
Much research effort has been devoted to the problem of finding
time-independent and time-dependent constants of motion
 for hamiltonian and nonhamiltonian ordinary differential equations.
 The reason is that except for some simple cases, this problem is very hard 
and no general methods to solve it are known up to now. Nevertheless several
 approaches were developed in the last years, thanks to the use of computer 
algebra facilities. The most important among them being: specific ansatz for a 
invariant [13-16], singularity analysis [17,18], the Lie symmetry method [19,20], 
the linear compatibility analysis method [21,22], rescaling 
method [23,24] and the Darboux method [25-28]. Among the first mentioned 
methods one can cite the Carleman embedding procedure [13], the generalised 
Carleman [14] and the Hamiltonian methods [15,16]. Although they differ
in the details of computation, all these methods are based on an a priori 
hypothesis on the form of the invariant or the first integral $H$.

In 1878 Darboux [25] showed how can be constructed the first integrals of 
planar polynomial ordinary differential equations possessing sufficient invariant 
algebraic curves. The Darboux method is based on the possibility of writing the 
invariant ( or at least an integrating factor) as the 
product of different algebraic functions $f_i$ raised at a given power $\lambda_i$.
It is on the form of these functions $f_i$ that it is introduced an Ansatz. The $f_i$ 
are determined straightforwardly identifying to zero the coefficients of a polynomial
expression in $x,y,z$ . In that sense the Darboux method is 
not so different from some of the others cited above. Nevertheless the experiance shows
that we have somehow divided the difficulties of the unavoidable identification. 
Typically the number of equations obtained is greater than the number of unknowns.
In order to satisfy the whole system of algebraic equations we must then introduce some {\it conditions} on the
parameters describing the system (\ref{l1}). From a physical point of view the 
rule of the game is to find first integrals with as few conditions as possible on 
the parameters of the given system.

Although the Darboux theory of integrability works for real or complex polynomial ordinary differential equations,
we are concerned here with the existence of real first integrals of (\ref{l1}) 
when the parameters $a_i,b_{ij}$ of the system, the three dependent variables $x$, $y$, $z$, and the independent variable $t$ (the {\it time}) are real.

The paper is organized as follows. The main lines of the Darboux theory for three-dimensional 
polynomial differential systems are presented in Section 2. The first integrals constructed 
using the Darboux theory are build here exclusively with polynomial invariant algebraic solutions.
In Section 3, the Proposition 1 exhibits the linear algebraic solutions of the LV3. 
The following sections contain the cases for which at least a first integral has been found.  
The cases where the LV3 has a single first integral are considered first, see 
Sections 4 and 5. In Section 4, the Theorems 2 and 3 concern the cases where all the $f_i$ 
are linear and in Section 5, the Theorems 4 and 5 are for the cases where $f_4$ is a 
quadratic polynomial, the first being when the conic passes through the origin and the second
when not. In Section 6 we consider the cases of integrability, i.e. the cases where two first integrals
coexist. Theorems 6--10 are when at most one of the first integrals contain quadratic
algebraic solutions and Theorem 11 is for the cases where the two first integrals are formed with a 
quadratic algebraic solution. Finally in Section 7 we give our conclusions.

\section{Computational method}

The Darboux method is based on the existence of algebraic invariant solutions. Suppose that
we can determine two polynomials $f_i(x,y,z)$ and $K_i(x,y,z)$ such that
\bq
\frac{\partial f_i}{\partial x}P_1+\frac{\partial f_i}{\partial y}P_2+\frac{\partial f_i}{\partial z}P_3=K_if_i\ .
\label{l3}
\eq 
Then equation $f_i(x,y,z)=0$ describes a surface which is formed by
 trajectories. It is an {\it algebraic solution} of the system. The polynomial $K_i$ is called the
{\it cofactor} of $f_i$. 

These $f_i$ are going to be the ``bricks'' with which we will build the invariants.
Suppose that we have obtained $q$ functions $f_i$. Let us consider the following function of the variables $x,y,z$ and $t$
\bq
H=\prod_{i=1}^q f_i^{\lambda_i}(x,y,z) \exp(st)\ .\label{l4}
\eq
Taking the total derivative
\bq
\begin{array}[b]{@{}l@{}}
\ds\frac{\mbox{d}H}{\mbox{d}t}=\frac{\partial H}{\partial t}+\frac{\partial H}{\partial x}P_1+
\frac{\partial H}{\partial y}P_2+\frac{\partial H}{\partial z}P_3\\[3ex]
\ds\qquad=H\left[s+\sum_{i=1}^q\frac{\lambda_i}{f_i}
\left(\frac{\partial f_i}{\partial x}P_1+\frac{\partial f_i}{\partial y}P_2+
\frac{\partial f_i}{\partial z}P_3\right)\right]\ .\label{l5}
\end{array}
\eq 

\noindent Imposing that $H$ is an invariant and making use of (\ref{l3}), we obtain
\bq
s+\sum_{i=1}^q \lambda_i K_i=0\ .\label{l6}
\eq 
The equations in the $\lambda_i$ are now linear equations.
How many $f_i$ do we need? The equation (\ref{l3}) shows that if the
system is of degree $m$, $K_i$ is at most of degree $m-1$ independently 
of the degree of $f_i$. So, the left hand side of (\ref{l6}) is a 
polynomial in $x,y,z$ of degree at most $m-1$ with a total 
of $(m+2)(m+1)m/6$ terms. Consequently (\ref{l6}) produces $(m+2)(m+1)m/6$ 
equations where the unknowns are the $\lambda_i.$ We see from 
(\ref{l6}) that if we allow $s$ to be different from zero the system 
of equations in the $\lambda_i$ is inhomogeneous, and we need 
$q=(m+2)(m+1)m/6.$ If we want a first integral, then $s=0$  and
the system becomes homogeneous and we need a priori either a new $f_i$,
 or a new condition by imposing that the determinant of the system of 
equations in the $\lambda_i$ is zero. This will be the case
here where we are concerned with first integrals exclusively. The required
number of $\lambda_i$ in order to have first integrals is then $q=(m+2)(m+1)m/6+1$. 
However, if it exists a solution with $q=(m+2)(m+1)m/6$, then
a first integral exists. In fact, all the first integrals presented here are with
this value of $q$ (Theorem 3 is for first integrals having a quadratic 
algebraic solutions which factorize to linear ones).  
We see that the possibility of solving for the $\lambda_i$ in the system deduced
identifying to zero the coefficients of the polynomial on $x,y,z$ deduced from equation (\ref{l6})
depends on the conditions we put on the coefficients of (\ref{l1}). 
We should not forget that some conditions
come from equation (\ref{l6}) and others from equation (\ref{l3}).

We make a comment on the number of conditions on the system parameters
in order to obtain a first integral and its variation with  the degree of 
the invariant solutions tested. For the existence of invariant solutions
of higher degree we require a larger number of conditions than for solutions of smaller
degree. At least this is what happens for degrees one to three. For higher degrees one can get 
a saturation as this is what happens in 2d (see Ref 29). It turns out that the most interesting 
cases (i.e. the cases with a not too high number of conditions) will
be obtained using straight lines and conics as invariant solutions for Lotka--Volterra
and quadratic systems.

In the case of system (\ref{l1}), $m=2$ and in this paper we take $q=4$. As
the axes are algebraic solutions, the problem of invariant search is reduced 
to the determination of only one algebraic
solution i.e. $f_4=0$. Moreover $f_4$ is taken here as polynomial of degree at
most two, namely $f_4\equiv(\overrightarrow{\nu} \cdot \overrightarrow{f})$
where $\overrightarrow{\nu}$ is a vector of dimension $10$ and with components 
$1$ or $0$ and $\overrightarrow{f}$ a vector of components 
$(f_{000}, f_{100}x, f_{010}y, f_{001}z, f_{200}x^2, f_{110}xy, f_{101}xz, 
f_{020}y^2, f_{011}yz,$ $f_{002}z^2)$.

The results presented here are given modulo the {\it three-dimensional Lotka-Volterra
 system equivalences}. This is because we can associate to a given LV3
(\ref{l1}) five
equivalent three-dimensional Lotka-Volterra systems. The first two are
obtained doing circular permutation of the variables $x,y,z$ and of the parameters 
$a_i$ and $b_{ij}$,  next  three systems $\forall b_{ij}\neq 0$
are obtained doing the transformation
\[
\begin{array}{@{}l@{}}
(x,\,\,\, y,\,\,\,\,\, z,\,\,\,\,\, a_1,\,\,\,\,\, a_2,\,\,\,\,\, a_3,\,\,\,\,\, 
b_{11},\,\,\,\,\, b_{12},\,\,\,\,\, b_{13},\,\,\,\,\, b_{21},\,\,\,\,\, b_{22},\,\,\,\,\, 
b_{23},\,\,\,\,\, b_{31},\,\,\,\,\, b_{32},\,\,\,\,\, b_{33}) \to\\[1ex]
\qquad(x,\,\,\,\,\, z,\,\,\,\,\, y,\,\,\,\,\, a_1,\,\,\,\,\, a_3,\,\,\,\,\, a_2,\,\,\,\,\, 
b_{11},\,\,\,\,\, b_{13},\,\,\,\,\, b_{12},\,\,\,\,\, b_{31},\,\,\,\,\, b_{33},\,\,\,\,\, 
b_{32},\,\,\,\,\, b_{21},\,\,\,\,\, b_{23},\,\,\,\,\, b_{22}),
\end{array}
\]
which keeps invariant the system, and the two others obtained doing circular
permutation of variables and parameters. We say that all these Lotka--Volterra 
systems are {\it $E$ equivalent}. All the results of this paper are stated modulo 
these {\it $E$ equivalences}.

\begin{table}[htbp]
\vspace{-1em}
\caption{Definition of the terminology used for the first integral conditions.}
\vspace{1em}
\small
\centering
\begin{tabular}{|c|l|c|l|}
\hline  
na. & definition & na. & definition\\
\hline
$a_i$ & $a_i$                & $i_{ij}$  &  $b_{ik}-b_{jk}-b_{kk}$\\
$c_{ij}$ & $a_i-a_j$         & $i_{ij}'$ &  $b_{ik}+b_{jk}-b_{kk}$\\
$c_{ij}'$ & $a_i-2a_j$       & $i_{ij}''$ &  $2b_{ik}-b_{jk}-2b_{kk}$\\
$c_{ij}''$ & $a_i+a_j$       & $i_{ij}'''$ &  $2b_{ik}-b_{jk}-b_{kk}$ \\ 
$c_{ij}'''$ & $a_i+2a_j$     & $j_{ij}$ &  $b_{ik}+b_{jk}-2b_{kk}$\\
$v_{ijk}$ & $a_i+a_j+a_k$    & $j_{ij}'$ & $b_{ik}+b_{jk}-4b_{kk}$\\
$v_{ijk}'$ & $a_i+a_j-a_k$   & $j_{ij}''$ &  $2b_{ik}+b_{jk}-4b_{kk}$\\ 
$v_{ijk}''$ & $a_i+a_j-2a_k$ & $n_{ij}$ &  $b_{ii}(b_{ij}-b_{jj})-(b_{ij}-2b_{jj})(b_{ji}-b_{ii})$\\ 
$v_{ijk}'''$ & $a_i+a_j+2a_k$ & $n_{ij}'$ &  $(b_{ij}-b_{jj})(b_{ki}-b_{ii})+(b_{ij}+b_{kj}-2b_{jj})(b_{ji}-b_{ii})$\\ 
$w_{ijk}$ & $2a_i+a_j+a_k$ & $n_{ij}''$ &  $b_{ii}(b_{ij}+b_{kj}-3b_{jj})-b_{ki}(b_{ij}-b_{jj})$\\
$w_{ijk}'$ & $2a_i+a_j-a_k$ & $n_{ij}'''$ &  $3b_{ii}b_{jj}+b_{ji}b_{kj}-b_{ii}(b_{ij}+b_{kj})-b_{jj}(b_{ji}+b_{ki})$\\
$w_{ijk}''$ & $2a_i-a_j-a_k$ & $n_{ij}^{iv}$ & $(b_{ij}-b_{jj})(b_{ki}-2b_{ii})-b_{jj}(b_{ji}-b_{ii})$\\
$w_{ijk}'''$ & $a_i+2(a_j+a_k)$ & $o_{ijk}$ &  $(b_{ij}-b_{jj})(b_{jk}-b_{kk})(b_{ki}-b_{ii})+(b_{ji}-b_{ii})(b_{kj}-b_{jj})(b_{ik}-b_{kk})$\\
$b_i$        & $b_i$            & $p_{ij}$ &  $2b_{ii}(b_{kj}-b_{jj})+b_{jj}(b_{ji}-b_{ii})$\\
$d_{ij} $ & $b_{ij}-b_{jj}$ & $p_{ij}'$ &  $2b_{ii}b_{jj}-b_{ij}(b_{ki}+b_{ii})$\\
$d_{ij}'$ & $2b_{ij}-b_{jj}$  & $p_{ij}''$ & $b_{ki}b_{kj}-2b_{ij}(2b_{ki}-b_{ji})$\\
$d_{ij}''$ & $b_{ij}-2b_{jj}$ & $p_{ij}'''$ & $b_{ii}b_{kj}-b_{ki}(b_{kj}-b_{jj})$\\
$e_{ij} $ & $b_{ij}+b_{jj}$ & $p_{ij}^{iv}$ &  $2b_{ii}(b_{ij}-b_{jj})+b_{jj}(b_{ki}-b_{ii})$\\
$e_{ij}'$ & $2b_{ij}+b_{jj}$ & $q_{ij}$ & $b_{ij}b_{ii}b_{kk}+b_{ji}b_{jj}(b_{ik}-2b_{kk})$\\
$e_{ij}''$ & $b_{ij}+2b_{jj}$ & $q_{ij}'$ & $b_{ii}b_{ik}(b_{ij}-b_{jj})+b_{jj}b_{jk}(b_{ji}-b_{ii})$\\
$g_{ij}$ &  $b_{ik}-b_{jk}$ & $q_{ij}''$ &  $b_{ii}b_{jj}(b_{kk}+b_{jk})+2b_{ki}(b_{ij}-2b_{jj})$\\
$g_{ij}'$ &  $b_{ik}-2b_{jk}$ & $q_{ij}'''$ &  $b_{ii}b_{jj}(b_{ki}-2b_{ii})-(b_{kj}-2b_{jj})(b_{ji}-2b_{ii})$\\
$g_{ij}''$ &  $b_{ik}+b_{jk}$ & $r_{ij}$ &  $a_ib_{jj}(b_{ji}-b_{ii})+a_jb_{ii}(b_{ij}-b_{jj})$\\
$g_{ij}'''$ &  $b_{ik}+2b_{jk}$ & $r_{ij}'$ &  $a_jb_{ij} - (a_i+2a_j)b_{jj}$\\
$h_{ij}$ &  $b_{ii}b_{kj}-b_{jj}b_{ki}$ & $r_{ij}''$ &  $a_jb_{ij} - (a_i+a_j)b_{jj}$\\
$h_{ij}'$ &  $b_{ii}b_{kj}-b_{jj}b_{ji}$  & $r_{ij}'''$ &  $2a_ib_{ji} - (a_i+a_j)b_{ii}$\\
$k_{ij} $ & $a_j b_{ij} - a_i b_{jj}$  & $r_{ij}^{iv} $ & $a_i b_{jk} - a_j (b_{ik}+2b_{jk})$\\
$k_{ij}' $ & $a_j b_{ij} + a_i b_{jj}$ & $s_{ij}$ &  $a_jb_{ij} - (a_i+a_j+a_k)b_{jj}$\\
$l_{ij} $ & $a_i b_{jk} - a_j b_{ik}$ & $s_{ij}'$ &  $2a_jb_{ij} - (a_i+2a_j+a_k)b_{jj}$\\
$l_{ij}' $ & $a_i b_{jk} + a_j b_{ik}$  & $s_{ij}''$ &  $(a_i+2a_j)b_{jj}-a_j(b_{ij}+b_{kj})$\\
$m_{ij} $ & $a_j b_{ij} + a_k b_{jj}$ & $s_{ij}'''$ & $(a_ib_{ji}-a_jb_{ii})(b_{kj}-b_{jj})(b_{jk}-b_{kk})+a_ib_{ii}(b_{jk}b_{kj}-b_{jj}b_{kk})$\\
\hline
\end{tabular}
\end{table}

\section{Invariant algebraic solutions}

In this section we study the invariant algebraic solutions of the LV3 systems of 
degree at most $1$. Thus, in Proposition 1, we present the invariant planes 
(invariant algebraic solutions of degree $1$) together with their cofactors, 
and the conditions for their existence.

\bigskip
\noindent{\bf Proposition 1.} {\it An LV3 has an invariant plane $f=0$ with cofactor 
$K$ in the following cases, modulo the $E$ equivalences, using the notation defined in Table 1
\begin{itemize}
\item[(1)]  $f=x=0$ with $K=b_{11}x+b_{12}y+b_{13}z$.
 
\item[(2)] If $b_{12}=b_{13}=0$ and $a_1b_{11}\neq 0$, then $f=a_1+b_{11}x=0$  and $K=b_{11}x$. 

\item[(3)] If $a_1=a_2$, $b_{13}=b_{23}$ and $d_{12}d_{21}\neq 0$ then $f=(b_{21}-b_{11})x-(b_{12}-b_{22})y=0$ 
 and $K=a_1+b_{11}x+b_{22}y+b_{13}z$.

\item[(4)] If $a_1=a_2=a_3$, $o_{123}=0$ and $d_{12}d_{21}d_{13}d_{31}\neq 0$, then 
$f=(b_{21}-b_{11})(b_{31}-b_{11})x-(b_{12}-b_{22})(b_{31}-b_{11})y-(b_{13}-b_{33})(b_{21}-b_{11})z=0$  
and $K=a_1+b_{11}x+b_{22}y+b_{33}z$.

\item[(5)] If $r_{12}=r_{23}=r_{31}=0$ and $a_1a_2a_3b_{11}b_{22}b_{33}\neq 0$, 
then $f=a_1a_2a_3+a_2a_3b_{11}x+a_1a_3b_{22}y+a_1a_2b_{33}z=0$ and $K=b_{11}x+b_{22}y+b_{33}z$.

\item[(6)] If $b_{13}=b_{23}=r_{12}=0$ and $a_1a_2b_{11}b_{22}\neq 0$, 
then $f=a_1a_2+a_2b_{11}x+a_1b_{22}y=0$ and $K=b_{11}x+b_{22}y$.

\item[(7)] If $a_3=b_{33}=r_{12}=o_{123}=0$ and $a_1a_2b_{11}b_{22}\neq 0$,
then $f=(b_{31}-b_{11})(a_1a_2+a_2b_{11}x-a_1b_{22}y)-a_2b_{11}b_{13}z=0$ and $K=b_{11}x+b_{22}y$.

\item[(8)] If $a_2=b_{22}=b_{13}=b_{23}=0$ and $a_1b_{11}b_{12}d_{21}\neq 0$,
then $f=(b_{21}-b_{11})(a_1+b_{11}x)-b_{11}b_{12}y=0$ and $K=b_{11}x$.

\item[(9)] If $a_2=a_3=b_{22}=b_{33}=o_{123}=0$ and $a_1b_{11}b_{32}d_{21}\neq 0$,
then $f=b_{32}(b_{21}-b_{11})(a_1+b_{11}x)-b_{11}b_{12}(b_{32}y-b_{23}z)=0$ and $K=b_{11}x$.

\item[(10)] If $a_2=a_3=b_{22}=b_{33}=b_{23}=b_{32}=0$ and $a_1b_{11}d_{21}d_{31}\neq 0$,
then $f=(b_{21}-b_{11})(b_{31}-b_{11})(a_1+b_{11}x)+b_{11}[b_{12}(b_{11}-b_{31})y+
b_{13}(b_{11}-b_{21})z]=0$ and $K=b_{11}x$.

\item[(11)] If $a_2=a_3=b_{22}=b_{33}=b_{12}=b_{32}=0$ and $a_1b_{11}d_{31}\neq 0$,
then $f=(b_{31}-b_{11})(a_1+b_{11}x)-b_{11}b_{13}z=0$ and $K=b_{11}x$.

\item[(12)] If $a_2=a_3=b_{22}=b_{33}=b_{12}=d_{21}=0$ and $a_1b_{11}b_{23}b_{32}d_{31}\neq 0$,
then $f=b_{23}(b_{31}-b_{11})(a_1+b_{11}x)+b_{11}b_{13}(b_{32}y-b_{23}z)=0$ and $K=b_{11}x$.

\end{itemize}}

\begin{proof}
The proof is obtained finding both the linear $f$ and $K$ satisfying 
equation (\ref{l3}). So doing, we introduce in (\ref{l3}) the general form for $f$ and $K$,
namely $f=f_{000}+f_{100}x+f_{010}y+f_{001}z$, $K=k_{000}+k_{100}x+k_{010}y+k_{001}z$. 
The problem consits then in the evaluation of the coefficients of $f$ and $K$ by solving
the algebraic system obtained identifying to zero the coefficients of the polynomial
in $x,y,z$ which results after this introduction of $f$ and $K$ in (\ref{l3}).
The number of equations of this algebraic system is usely greater than the number of 
the unknown coefficients of $f$ and $K$. In order to satisfy the entire system of
algebraic equations, it is necessary then to introduce the conditions on the parameters
of the differential system (\ref{l1}), which appear in the different statements of the theorem.
\end{proof}

\newpage
\section{First integrals formed with linear invariant algebraic solutions}

In this section we apply the Darboux method to deduce first 
integrals obtained with linear polynomial algebraic solutions.

\bigskip
\noindent{\bf Theorem 2.} {\it Let $f_1=x$, $f_2=y$ and $f_3=z$. If a LV3 has a
 fourth algebraic solution $f_4=0$ of degree 1, then using the 
notation defined in Table 1, we can establish the following statements 

\smallskip 
\noindent $(1)$ If the LV3 satisfy to the conditions $\det\,(b_{ij})=0$ and $s=\sum_{i=1}^{3} a_i\a_i=0$, 
where the $\a_i$ are any non-trivial solution of $(b_{ij})^T (\a_i) =0$, then the function $|x|^{\a_1}|y|^{\a_2}|z|^{\a_3}$  
is a first integral. 

\smallskip 
\noindent $(2)$ If the conditions $c_{12}=c_{23}=o_{123}=0$  are satisfied and 
$d_{12}d_{21}d_{13}d_{31}\neq 0$, then the  plane $f_4 \equiv \phi=(b_{21}-b_{11})(b_{31}-b_{11})x-(b_{12}-b_{22})
(b_{31}-b_{11})y-(b_{13}-b_{33})(b_{21}-b_{11})z=0$
is an algebraic solution of the LV3  and a first integral $|f_1|^{\a_1}|f_2|^{\a_2}|f_3|^{\a_3}f_4$
exists, the $\a_i$ being defined as solution of the system  $(b_{ij})^T (\a_i) = -\,\mbox{diag} (b_{ij}).$ 

\smallskip 
\noindent $(3)$ If the conditions $c_{12}=g_{12}=k_{13}=0$ 
are satisfied and $d_{12}d_{21}(|b_{11}|+|b_{21}|+|b_{31}|)\neq 0$, then  the plane $f_4=
(b_{21}-b_{11})x+(b_{22}-b_{12})y=0,$ is an algebraic solution of the LV3 and a first integral 
$|f_1|^{\la_1}|f_2|^{\la_2}|f_3|^{\la_3}|f_4|^{\la_4}$ exists with  $\la_1=(b_{21}-b_{11})
(b_{13}b_{32}-b_{33}b_{22}),\la_2=(b_{12}-b_{22})(b_{13}b_{31}-b_{33}b_{11}),
\la_3=b_{13}(b_{12}-b_{22})(b_{11}-b_{21}),
\la_4=b_{13}[b_{32}(b_{11}-b_{21})+b_{31}(b_{22}-b_{12})]+b_{33}(b_{12}b_{21}-b_{11}b_{22})$.

\smallskip 
\noindent $(4)$ If the conditions $b_{13}=b_{23}=r_{12}=0$ 
are satisfied and $a_1a_2b_{11}b_{22}d_{12}d_{21}\neq 0,$ then,  the plane $f_4=
a_1a_2+a_2b_{11}x+a_1b_{22}y=0,$ is an algebraic solution of the LV3 and a first integral 
$|f_1|^{\la_1}|f_2|^{\la_2}f_3^{\la_3}|f_4|^{\la_4}$ exists with $\la_1=b_{22}(b_{21}-b_{11}), 
\la_2=b_{11}(b_{12}-b_{22}), \la_3=0, \la_4=b_{11}b_{22}-b_{12}b_{21}$.  

\smallskip 
\noindent $(5)$ If the conditions $r_{12}=r_{23}=r_{31}=a_1a_3b_{11}b_{22}d_{23}+ 
a_1a_2b_{22}b_{33}d_{31}+a_2a_3b_{11}b_{33}d_{12}=0$  are satisfied and 
$a_1a_2a_3b_{11}b_{22}b_{33}\neq 0$, then the  plane $f_4=a_1a_2a_3+a_2a_3b_{11}x+a_1a_3b_{22}y+a_1a_2b_{33}z=0$
is an algebraic solution of the LV3  and a first integral $|f_1|^{\a_1}|f_2|^{\a_2}|f_3|^{\a_3}f_4$
exists, the $\a_i$ being defined as solution of the system  $(b_{ij})^T (\a_i) = -\,\mbox{diag} (b_{ij}).$ 

\smallskip 
\noindent $(6)$ If the conditions $b_{13}=b_{23}=d_{12}=d_{21}=0$ 
are satisfied and $a_1a_2c_{12}\neq 0,$ then,  
the plane $f_4=a_1a_2+a_2b_{11}x+a_1b_{22}y=0,$ is an algebraic solution of the 
LV3 and a first integral $|f_1|^{\la_1}|f_2|^{\la_2}f_3^{\la_3}|f_4|^{\la_4}$ exists 
with $\la_1=a_2, \la_2=-a_1, \la_3=0, \la_4=a_1-a_2$.  

\smallskip 
\noindent $(7)$ If the conditions $a_2=b_{13}=b_{22}=b_{23}=0$ and 
$b_{11}b_{12}b_{21}d_{21}\neq 0$, then  the plane $f_4=(b_{21}-b_{11})
(a_1+b_{11}x)-b_{11}b_{12}y=0,$ is an algebraic solution of 
the LV3 and a first integral $f_1^{\la_1}|f_2|^{\la_2}f_3^{\la_3}|f_4|^{\la_4}$ 
exists with $\la_1=0, \la_2=-b_{11}, \la_3=0, \la_4=b_{21}.$

\smallskip 
\noindent $(8)$ If the conditions $b_{12}=b_{13}=b_{22}=b_{23}=0$  
are satisfied and $a_1b_{11}k_{21}\neq 0$, then  the plane 
$f_4=a_1+b_{11}x=0,$ is an algebraic solution of the LV3 and a first integral 
$|f_1|^{\la_1}|f_2|^{\la_2}f_3^{\la_3}|f_4|^{\la_4}$ exists with $\la_1=a_2b_{11}, 
\la_2=-a_1b_{11}, \la_3=0, \la_4=a_1b_{21}-a_2b_{11}$.  

\smallskip 
\noindent $(9)$ If the conditions $c_{32}'=b_{12}=b_{13}=d_{23}'=d_{32}''=i_{23}'''=0$ 
are satisfied and $a_1b_{11}\neq 0$, then  the plane $f_4=a_1+b_{11}x=0,$ is an 
algebraic solution of the LV3 and a first integral $f_1^{\la_1}f_2^{\la_2}f_3^{\la_3}
f_4^{\la_4}$ exists with $\la_1=0, \la_2=-2,\la_3=1, \la_4=1.$

\smallskip 
\noindent $(10)$ If the conditions $v_{231}'=b_{12}=b_{13}=e_{23}=e_{32}=j_{23}=0$ 
are satisfied and $a_1b_{11}\neq 0$, then  the plane $f_4=a_1+b_{11}x=0,$ is an 
algebraic solution of the LV3 and a first integral $f_1^{\la_1}f_2^{\la_2}f_3^{\la_3}
f_4^{\la_4}$ exists with $\la_1=1, \la_2=-1, \la_3=-1, \la_4=1.$

\smallskip 
\noindent $(11)$ If the conditions $c_{23}''=b_{12}=b_{13}=e_{23}=e_{32}=j_{23}=
0$ are satisfied and $a_1b_{11}\neq 0$, then the plane $f_4=a_1+b_{11}x=0,$ 
is an algebraic solution of the LV3 and a first integral $f_1^{\la_1}f_2^{\la_2}
f_3^{\la_3}f_4^{\la_4}$ exists with $\la_1=0, \la_2=-1, \la_3=-1, \la_4=2.$

\smallskip 
\noindent $(12)$ If the conditions $c_{23}=b_{12}=b_{13}=b_{23}=b_{33}=d_{31}=d_{32}=0$
 are satisfied and $a_1b_{11}d_{21}\neq 0$, then the plane $f_4=a_1+b_{11}x=0,$ is an algebraic 
solution of the LV3 and a first integral $f_1^{\la_1}|f_2|^{\la_2}|f_3|^{\la_3}|f_4|^{\la_4}$ 
exists with $\la_1=0, \la_2=b_{11}, \la_3=-b_{11}, \la_4=b_{11}-b_{21}.$

\smallskip 
\noindent $(13)$ If the conditions $c_{31}=c_{21}'=b_{12}=b_{13}=d_{21}=d_{31}=
d_{32}'=d_{23}''=0$ are satisfied and $a_1b_{11}\neq 0$, then  the plane 
$f_4=a_1+ b_{11}x=0,$ is an algebraic solution of the LV3 and a first integral 
$f_1^{\la_1}f_2^{\la_2}f_3^{\la_3}f_4^{\la_4}$ exists with 
$\la_1=0, \la_2=1, \la_3=-2, \la_4=1.$

\smallskip 
\noindent $(14)$ If the conditions $c_{12}=c_{31}''=b_{13}=b_{23}=b_{33}=e_{31}=e_{32}=0$ 
are satisfied and $d_{12}d_{21}\neq 0$, then 
 the plane $f_4=(b_{21}-b_{11})x -(b_{12} - b_{22})y=0,$ is an algebraic solution of 
the LV3 and a first integral $f_1^{\la_1}f_2^{\la_2}f_3^{\la_3}f_4^{\la_4}$ exists with 
$\la_1=0, \la_2=0, \la_3=1, \la_4=1.$

\smallskip 
\noindent $(15)$ If the conditions $c_{12}=c_{31}'=b_{11}=b_{13}=b_{22}=b_{23}=
b_{31}=b_{33}=g_{13}=0$ are satisfied and $b_{12}b_{21}\neq 0$, then  the plane 
$f_4=b_{21}x - b_{12}y=0,$ is an algebraic solution of the LV3 and a first integral 
$f_1^{\la_1}f_2^{\la_2}f_3^{\la_3}f_4^{\la_4}$ exists with 
$\la_1=1, \la_2=0, \la_3=-1, \la_4=1.$
}
\begin{proof}
The cofactors of $f_1, f_2, f_3$ are 
$K_1=a_1+b_{11}x+b_{12}y+b_{13}z$, $K_2=a_2+b_{21}x+
b_{22}y+b_{23}z$, $K_3=a_3+b_{31}x+b_{32}y+b_{33}z$ 
respectively. Statement $(1)$ use only these three algebraic solutions. 
For the other statements, it is easy to check, that under suitable assumptions, 
$f_4=0$ is an algebraic solution of the LV3, i.e. that $f_4$ verifies 
eq (\ref{l3}) where $K_4$, the cofactor of $f_4$  takes the values
$K_4=0$ for statement $(1)$ where $f_4=1$,
$K_4= a_1 + b_{11} x + b_{22} y + b_{33} z $ for statement $(2)$,
$K_4= a_1 + b_{11} x + b_{22} y + b_{13} z $ for statements $(3), (14)$ and $(15)$,
$K_4= b_{11} x + b_{22} y $ for statement $(4)$, 
$K_4= b_{11} x + b_{22} y + b_{33} z $ for statement $(5)$, 
$K_4= b_{11} x $ for statements $(7)-(13)$. We note that $f_4$ is the algebraic
solution of Proposition 1(1) for statement $(1)$, of Proposition 1(4) for statement $(2)$,
of Proposition 1(3) for statements $(3), (14)$ and $(15)$
of Proposition 1(6) for statements $(4)$ and (6),
of Proposition 1(5) for statement $(5)$,
of Proposition 1(8) for statement $(7)$,
of Proposition 1(2) for statements $(8)-(13)$.
 Hence to each statement corresponds a first integral $|x|^{\la_1}|y|^{\la_2}|z|^{\la_3}
|f_4|^{\la_4}$.
\end{proof}

We must mention that statements (1)--(5) are, respectively, the 
invariants I, II, II', III' and III of Cair\'o and Feix [14] obtained using the
Carleman method (with the additional condition $s=0$ for statements (1), (3) and (5).
Note that statements (2) and (3) with $a_1=a_2=a_3=0$ concerns the ABC system when 
$b_{11}=b_{22}=b_{33}=0$ (see for instance Ref. 22). With these additional conditions 
statement (2) concerns an integrable system (Theorem 6(22)).

\bigskip
\noindent{\bf Theorem 3.} {\it The LV3 has a first integral $|f_1|^{\la_1}|f_2|^{\la_2}|f_3|^{\la_3}
|f_4|^{\la_4}|f_5|^{\la_4}$ formed with the coordinate axes $f_1=x$, $f_2=y$ and $f_3=z$ and two other 
algebraic solutions $f_4=0$ and $f_5=0$ of degree 1 in the cases described by the following statements 
 
\smallskip 
\noindent $(1)$ If the conditions $c_{23}=b_{11}=b_{21}=b_{23}=b_{31}=0$  are satisfied 
 and $c_{13}b_{22}b_{33}(a_1b_{22}b_{33}+a_3(b_{13}(b_{32}-b_{22})-b_{12}b_{33}))\neq 0$,  
then  $f_4 =a_3+b_{22}y$ and $f_5=(b_{32}-b_{22})y+b_{33}z$  and
$\la_1= a_3b_{33}(b_{32}-2b_{22}),$
$\la_2= (b_{32}-2b_{22})(a_3b_{13}-a_1b_{33}),$
$\la_3= a_1b_{22}b_{33}-a_3(b_{12}b_{33}-b_{13}b_{22})$ and
$\la_4= -a_1b_{22}b_{33}+a_3(b_{12}b_{33}+b_{13}b_{22}-b_{13}b_{32}).$

\smallskip 
\noindent $(2)$ If the conditions $c_{23}=b_{12}=b_{13}=g_{23}=s_{12}'''=0$ are satisfied  and 
$a_1b_{11}d_{32}k_{21}\neq0$, then  $f_4 =a_1+b_{11}x$ and $f_5=(b_{32}-b_{22})y-(b_{23}-b_{33})z$
and $\la_1= a_2(b_{32}-b_{22})(b_{23}-b_{33}),$ $\la_2=a_1b_{33}(b_{32}-b_{22})$, $\la_3= a_1b_{22}(b_{23}-b_{33}),$
$\la_4=a_1(b_{22}b_{33}-b_{23}b_{32}).$

\smallskip 
\noindent $(3)$ If the conditions $b_{12}=b_{13}=r_{12}=r_{23}=r_{31}=r_{23}'''=0$ are satisfied  and 
$a_1a_2a_3c_{23}b_{11}b_{22}b_{33}\neq0$, then  $f_4 =a_1+b_{11}x$ and $f_5=a_1a_2a_3+a_2a_3b_{11}x+
a_1a_3b_{22}y+a_1a_2b_{33}z$ and  $\la_1=0$, $\la_2=2a_{3}/(a_{2}-a_{3})$, $\la_3=-2a_2/(a_2-a_3)$ and $\la_4=1$.

\smallskip 
\noindent $(4)$ If the conditions $c_{23}=g_{23}=h_{23}=p_{13}'=r_{12}=r_{23}=0$ are satisfied  and 
$b_{11}b_{12}b_{13}b_{22}b_{33}\neq0$, then  $f_4 =b_{22}y+b_{33}z$ and $f_5=(a_1+b_{11}x)(b_{21}+b_{11})+b_{11}b_{22}y+b_{11}b_{33}z$ 
and $\la_1=-(b_{21}+b_{11})/b_{11}$, $\la_2=\la_3=0$ and $\la_4=1$.

\smallskip 
\noindent $(5)$ If the conditions $c_{23}=b_{12}=b_{13}=b_{22}=d_{23}'=g_{23}=s_{21}'=0$ are satisfied  and 
$a_1b_{11}b_{23}b_{32}\neq0$, then  $f_4 =a_1+b_{11}x$ and $f_5=b_{32}y+b_{23}z$ 
and $\la_1=a_2/a_1$, $\la_2=-2$, $\la_3=0$ and $\la_4=1$.

\smallskip 
\noindent $(6)$ If the conditions $a_2=c_{31}''=b_{12}=b_{13}=b_{22}=b_{31}=d_{21}=d_{23}'=0$ are satisfied  and 
$a_1a_3b_{11}\neq 0$, then  $f_4 =a_1+b_{11}x$ and $f_5=a_1+b_{11}x+2b_{32}y+b_{33}z$ 
and $\la_1=\la_3=0$, $\la_2=-2$ and $\la_4=1$.

\smallskip 
\noindent $(7)$ If the conditions $a_2=a_3=b_{12}=b_{13}=b_{23}=b_{32}=d_{21}=g_{23}=0$ are satisfied  and 
$a_1b_{11}b_{22}b_{33}\neq 0$, then  $f_4 =a_1+b_{11}x$ and $f_5=b_{22}y-b_{33}z$ 
and $\la_1=0$, $\la_2=\la_3=-1$  and $\la_4=1$.

\smallskip 
\noindent $(8)$ If the conditions $a_2=a_3=b_{12}=b_{13}=b_{33}=d_{32}'=d_{21}=g_{23}=0$ are satisfied  and 
$a_1b_{11}b_{23}b_{32}\neq 0$, then  $f_4 =a_1+b_{11}x$ and $f_5=b_{32}y+b_{23}z$ 
and $\la_1=\la_2=0$, $\la_3=-2$  and $\la_4=1$.
}
\begin{proof}
The first integral can be written as $|f_1|^{\la_1}|f_2|^{\la_2}|f_3|^{\la_3}
|f|^{\la_4}$ with $f=f_4f_5$. The cofactors of $f_1, f_2, f_3$ are 
$K_1=a_1+b_{11}x+b_{12}y+b_{13}z$, $K_2=a_2+b_{21}x+b_{22}y+b_{23}z$, $K_3=a_3+b_{31}x+b_{32}y+b_{33}z$ 
respectively. Now it is easy to check that for statements (1), (2), (5)-(9) $f_4$ and $f_5$ are the 
algebraic solutions of Proposition 1 (2) and (3), for statement (3)  $f_4$ and $f_5$ are the 
algebraic solutions of Proposition 1 (2) and (5), for statement (4)  $f_4$ and $f_5$ are the 
algebraic solutions of Proposition 1 (3) and (5) and for statement (5)  $f_4$ and $f_5$ are the 
algebraic solutions of Proposition 1 (2) and (7). Consequently $f=0$ is an algebraic solution of the LV3, 
i.e.  $f$ verifies eq (\ref{l3}) where $K$, the cofactor of $f$  takes the value
$K= a_3+ 2 b_{11} x + 2 b_{22} y + 2 b_{33} z$ for statement $(1)$,
$K= a_3+(b_{21}+b_{11})x + b_{22} y + b_{33} z$ for statement $(2)$,
$K= 2 b_{11} x + b_{22} y + 2 b_{33} z$ for statement $(3)$,
$K= a_1(b_{21}+b_{11})/b_{11}+(b_{21}+b_{11})x + 2 b_{22} y + 2 b_{33} z$ for statement $(4)$,
$K= 2 b_{11} x + 2 b_{22} y$ for statement $(5)$,
$K= a_2 + b_{11} x + 2 b_{22} y$ for statement $(6)$,
$K= 2 b_{11} x + b_{22} y + b_{33} z$ for statement $(7)$ and
$K= 2 b_{11} x + b_{22} y$ for statement $(8)$.
For each statement, a solution of system $\la_1K_1+\la_2K_2+\la_3K_3+\la_4K=0$.
Hence to each statement corresponds a first integral 
$|x|^{\la_1}|y|^{\la_2}|z|^{\la_3}|f_4|^{\la_4}|f_5|^{\la_4}$.
\end{proof}

The next two theorems concern LV3 systems having one first integral alone of type  
$|x|^{\la_1}|y|^{\la_2}|z|^{\la_3}|f_4|^{\la_4}$ with $f_4$ quadratic. 
These cases are characterized by $n$ conditions among the parameters of the systems. 
The expression of all the corresponding first integrals are available in the e-mail 
address: ``lcairo@labomath.univ-orleans.fr''. To prove each statement one must simply apply 
(\ref{l3}) to obtain $f_4$ and its corresponding cofactors $K_4$ and compute the $\la_i$ from the 
equation $\la_1K_1+ \la_2K_2+ \la_3K_3+ \la_4K_4=0$, where $K_1,K_2,K_3$ are the cofactors 
above mentioned for the coordinate axes, i.e. $K_1=a_1+b_{11}x+b_{12}y+b_{13}z$, 
$K_2=a_2+b_{21}x+b_{22}y+b_{23}z$, $K_3=a_3+b_{31}x+b_{32}y+b_{33}z$.

\section{First integrals with $f_4$ quadratic}

\noindent{\bf Theorem 4.} {\it Let $f_1=x$, $f_2=y$ and $f_3=z$. If a LV3 has a
 fourth algebraic solution, $f_4=0$ of degree 2 passing through the origin, then 
a first integral $|f_1|^{\la_1}|f_2|^{\la_2}|f_3|^{\la_3}|f_4|^{\la_4}$  can be obtained 
when the conditions given in Table 2 are fulfilled.} 
{\small
\begin{longtable}{|c|c|l|l|l|}
  \caption{First integral conditions of Theorem 4.} \\
  \hline
  st. & $\overrightarrow{\nu}$ & n & conditions =0&  conditions  $\neq  0$\\
  \hline
  \endfirsthead
  \caption[]{First integral conditions of Theorem 4 (continued).} \\
  \hline
  st. & $\overrightarrow{\nu}$ & n & conditions =0&  conditions  $\neq  0$\\
  \hline
  \endhead
  \hline
  \endfoot
$(1)$ & $(0000011111)$ &  4  & $c_{12}=c_{31}=n_{23}'=g_{32}=0$ &$d_{12}d_{13}d_{23}j_{13}$\\
$(2)$ & $(0011000111)$ &  4  & $c_{23}=b_{31}=b_{21}=n_{23}=0$ &$b_{22}b_{33}d_{23}d_{23}''d_{32}d_{32}''$\\
$(3)$ & $(0001010010)$ &  5  & $v_{123}'=b_{23}=i_{32}=g_{31}=k_{21}=0$ & $(^1)$\\
$(4)$ & $(0001101001)$ &  5  & $c_{31}'=b_{12}=b_{22}=b_{32}=n_{31}=0$ & $(^2)$\\
$(5)$ & $(0000111111)$ &  5  & $a_1=a_2=a_3=n_{12}'=n_{23}'=0$ & $(^3)$\\
$(6)$ & $(0000001111)$ &  5  & $a_1=a_2=a_3=i_{23}'''=n_{23}'=0$ & $(^4)$\\
$(7)$ & $(0011001111)$ &  6  & $c_{32}=b_{21}=e_{31}=n_{32}'=r_{13}=r_{23}=0$ &$a_1a_2a_3c_{12}''b_{22}b_{33}$\\
$(8)$ & $(0011011001)$ &  6  & $c_{23}=b_{12}=b_{13}=d_{32}'=g_{23}=r_{31}=0$ &$b_{11}d_{23}''d_{21}$\\
$(9)$ & $(0011001100)$ &  6  & $c_{23}=b_{13}=b_{21}=d_{23}'=e_{31}=j_{31}=0$ &$a_1a_2a_3c_{12}b_{12}b_{33}$\\
(10)  & $(0011010110)$ &  6  &$c_{23}=b_{21}=b_{23}=d_{13}=d_{31}=r_{12}=0$ & $b_{13}b_{22}d_{12}$\\
$(11)$ & $(0011001110)$ &  6  & $c_{32}=b_{23}=b_{21}=b_{13}=e_{31}=s_{12}''=0$ &$a_1a_2a_3d_{32}$\\
$(12)$ & $(0011111111)$ &  6  & $c_{21}'=c_{23}=n_{12}=r_{23}=n_{31}=g_{32}=0$ &$b_{11}b_{22}b_{33}d_{23}d_{32}d_{31}''$\\
$(13)$ & $(0001111111)$ &  6  & $c_{12}=c_{32}'=d_{12}=n_{23}=n_{31}=d_{21}=0$ & $(^5)$\\
$(14)$ & $(0001111111)$ &  6  & $c_{12}=c_{32}'=n_{23}=n_{31}=n_{21}'=n_{13}'=0$ &$b_{12}b_{13}b_{33}d_{23}''d_{31}''i_{31}$\\
$(15)$ & $(0001011111)$ &  6  & $c_{12}=c_{32}'=o_{123}=n_{23}=r_{31}=g_{32}=0$ &$b_{21}b_{22}b_{33}d_{13}'d_{23}''$\\
$(16)$ & $(0011101001)$ &  6  & $c_{21}'=c_{23}=d_{12}'=g_{31}=r_{21}=n_{31}=0$ &$b_{33}d_{13}''d_{23}''g_{21}'$\\
$(17)$ & $(0011110100)$ &  6  & $c_{21}'=c_{23}=d_{31}''=d_{13}'=g_{21}=n_{12}=0$ &$b_{22}d_{21}''d_{32}''g_{31}'$\\
(18)   & $(0011111111)$ &  6  &$c_{23}=c_{21}'=n_{12}=n_{31}=o_{123}=n_{23}=0$ & $b_{22}b_{33}d_{32}d_{32}''d_{13}''$\\
$(19)$ & $(0001010111)$ &  6  & $c_{12}=c_{31}'=b_{31}=e_{21}=n_{23}=n_{32}'=0$ &$b_{12}b_{13}d_{12}i_{31}$\\
$(20)$ & $(0001011110)$ &  6  & $c_{12}=c_{31}'=b_{13}=b_{23}=g_{32}=n_{12}'''=0$ &$b_{11}b_{33}d_{21}d_{32}d_{32}''$\\
$(21)$ & $(0001111001)$ &  6  & $c_{12}=c_{31}'=b_{13}=b_{23}=g_{31}=n_{12}^{iv}=0$ &$b_{22}d_{21}d_{31}d_{31}''$\\
$(22)$ & $(0001110110)$ &  6  & $c_{12}=c_{31}'=b_{23}=d_{31}''=d_{13}'=n_{12}'=0$ &$b_{13}d_{32}d_{32}''g_{31}'$\\
$(23)$ & $(0001111100)$ &  6  & $c_{12}=c_{31}'=b_{13}=d_{23}'=n_{12}'=r_{32}=0$ &$b_{12}d_{31}d_{31}''g_{32}'$\\
$(24)$ & $(0111000011)$ &  6  & $c_{12}=c_{23}=b_{32}=d_{21}'=d_{31}'=n_{23}''=0$ &$a_1a_2a_3d_{13}''d_{23}$\\
$(25)$ & $(0001010110)$ &  6  & $c_{12}=c_{32}'=b_{23}=d_{12}=d_{21}=g_{32}'=0$ &$b_{13}b_{33}d_{13}d_{32}''$\\
$(26)$ & $(0001011100)$ &  6  & $c_{12}=c_{31}'=b_{13}=d_{23}'=d_{32}''=g_{32}=0$ &$b_{11}b_{12}b_{21}d_{21}$\\
$(27)$ & $(0011101000)$ &  6  & $c_{21}'=c_{23}=b_{13}=d_{12}'=d_{21}''=g_{31}=0$ &$b_{23}b_{33}d_{23}d_{31}''$\\
$(28)$ & $(0011111000)$ &  6  & $c_{21}'=c_{23}=b_{12}=b_{13}=g_{32}=r_{23}=0$ &$b_{22}b_{33}d_{31}''d_{32}$\\
(29)   & $(0011111000)$ &  6  &$c_{21}'=c_{23}=b_{12}=b_{13}=d_{32}=d_{23}=0$ & $d_{21}d_{21}''d_{31}d_{31}''g_{23}$\\
$(30)$ & $(0001001111)$ &  6  & $a_1=c_{32}'=b_{11}=n_{23}=n_{32}'=g_{32}'=0$ &$b_{12}b_{13}b_{33}d_{13}d_{32}''$\\
(31)   & $(0001000111)$ &  6  & $a_1=c_{32}'=b_{11}=b_{21}=b_{31}=n_{32}=0$ & $(^6)$\\
$(32)$ & $(0001010011)$ &  6  & $a_3=c_{12}''=b_{31}=b_{33}=e_{21}=g_{31}=0$ &$b_{12}b_{22}b_{23}d_{12}g_{12}''$\\
(33)   & $(0001000111)$ &  6  &$c_{12}=c_{32}'=b_{11}=b_{21}=b_{31}=n_{23}=0$ & $(^7)$\\
(34)   & $(0001101000)$ &  6  &$c_{32}=c_{21}'=b_{12}=b_{13}=b_{22}=b_{32}=0$ & $(^8)$\\
(35)   & $(0001111111)$ &  7  & $c_{12}=c_{32}'=b_{12}=b_{22}=d_{21}=d_{23}''=n_{31}=0$ & $b_{11}b_{33}b_{32}d_{13}''$\\
(36)   & $(0011010000)$ &  7  &$c_{23}=b_{12}=b_{31}=b_{32}=b_{33}=e_{21}=g_{21}''=0$ & $a_1a_2$\\
$(37)$ & $(0111111111)$ &  7  & $c_{12}=c_{23}=r_{23}'=r_{31}=r_{21}=r_{32}'=n_{13}'=0$ &$a_1a_2a_3b_{11}b_{22}b_{33}$\\
$(38)$ & $(0011111111)$ &  7  & $c_{23}=c_{21}'=n_{12}=r_{23}=n_{31}=n_{21}'=n_{13}'=0$ & $(^9)$\\
$(39)$ & $(0111001111)$ &  7  & $c_{12}=c_{13}=b_{31}=d_{21}'=n_{23}'=n_{13}''=r_{32}=0$ &$b_{13}b_{22}b_{32}e_{32}'j_{31}$\\
$(40)$ & $(0001011111)$ &  7  & $c_{12}=c_{32}'=b_{22}=o_{123}=n_{23}=r_{31}=g_{32}=0$ &$b_{32}b_{33}g_{31}'$\\
$(41)$ & $(0011101011)$ &  7  & $c_{21}'=c_{31}'=b_{32}=d_{12}'=n_{13}=n_{31}'=r_{21}=0$ &$b_{11}b_{12}b_{13}b_{23}b_{31}d_{31}''$\\
$(42)$ & $(0011111011)$ &  7  & $c_{21}'=c_{32}=b_{12}=b_{32}=o_{123}=n_{31}=i_{23}''=0$ &$b_{33}d_{13}''d_{21}''$\\
$(43)$ & $(0011111011)$ &  7  & $c_{21}'=c_{32}=b_{12}=b_{32}=g_{23}=n_{31}=n_{13}'=0$ &$b_{22}b_{31}b_{33}d_{31}''$\\
$(44)$ & $(0111001111)$ &  7  & $c_{12}=c_{13}=b_{31}=b_{32}=d_{21}'=n_{23}=n_{32}'=0$ &$b_{12}b_{13}b_{22}b_{33}d_{12}''$\\
(45)   & $(0001111111)$ &  7  & $c_{12}=c_{32}'=b_{11}=b_{21}=d_{12}=n_{23}=d_{13}''=0$ & $b_{31}b_{33}d_{23}''$\\
$(46)$ & $(0111001110)$ &  7  & $c_{12}=c_{32}=b_{23}=b_{31}=b_{13}=d_{21}'=j_{13}''=0$ &$a_1b_{11}b_{22}b_{32}b_{33}$\\
$(47)$ & $(0011111010)$ &  7  & $c_{21}'=c_{31}'=b_{12}=b_{23}=b_{32}=b_{13}=j_{23}'=0$ &$b_{22}d_{31}d_{21}''d_{31}''$\\
(48)   & $(0001000111)$ &  7  & $c_{12}=c_{32}'=b_{11}=b_{21}=b_{31}=b_{22}=d_{23}''=0$ & $(^{10})$\\
$(49)$ & $(0111011011)$ &  7  & $c_{13}=c_{12}=b_{12}=b_{31}=b_{21}=b_{32}=n_{23}''=0$ &$b_{11}b_{33}d_{23}$\\
(50)   & $(0010001010)$ &  7  & $a_1=c_{23}=b_{11}=b_{32}=d_{23}=d_{13}=g_{23}=0$ & $b_{22}d_{12}$\\
(51)   & $(0001011000)$ &  7  & $a_2=c_{13}=b_{13}=b_{22}=b_{33}=g_{32}=g_{31}=0$ & $b_{11}b_{23}d_{21}$\\
(52)   & $(0001000111)$ &  7  & $a_1=c_{32}'=b_{11}=b_{21}=b_{22}=b_{31}=d_{23}''=0$ & $(^{11})$\\
(53)   & $(0011001111)$ &  7  & $a_2=a_3=b_{21}=b_{22}=b_{33}=e_{31}=n_{23}'=0$ & $b_{11}b_{13}$\\
(54) & (0001101011) & 7 &$a_1=a_2=a_3=n_{13}=n_{13}'=q_{12}=q_{21}''=0$ & $(^{12})$\\
(55) & (0111001111) & 8 &$c_{31}=c_{21}=b_{13}=b_{31}=d_{21}'=e_{32}'=n_{12}'=n_{23}'=0$ & $b_{22}b_{33}$\\
(56) & (0011101000) & 8 &$c_{23}=c_{31}'=b_{13}=b_{33}=d_{31}=d_{12}'=g_{31}=g_{23}'=0$ & $a_1b_{11}b_{22}b_{23}$\\
(57) & (0011110100) & 8 &$c_{21}'=c_{23}=b_{13}=b_{23}=b_{33}=d_{31}''=d_{32}''=n_{12}=0$ & $b_{22}d_{21}''$\\
(58) & (0001111110) & 8 &$c_{12}=c_{32}'=b_{11}=b_{13}=b_{22}=b_{23}=g_{23}'=g_{13}'''=0$ & $b_{21}b_{33}$\\
(59) & (0011001100) & 8 &$a_1=c_{23}=b_{11}=b_{13}=b_{21}=b_{31}=d_{23}'=j_{13}=0$ & $a_2d_{32}''$\\
(60) & (0001101011) & 8 & $a_1=a_2=a_3=b_{11}=d_{13}''=n_{13}'=n_{23}'''=p_{23}''=0$ & $(^{13})$\\
(61) & (0001111100) & 8 &$a_1=a_2=a_3=b_{13}=d_{23}'=d_{31}''=d_{32}''=n_{12}'=0$ & $b_{11}b_{22}b_{12}b_{21}$\\
(62) & (0111010111) & 8 &$a_1=a_2=a_3=b_{21}=d_{31}'=j_{21}'=n_{23}=n_{23}'=0$ & $b_{22}b_{33}d_{13}''(d_{13}''-4b_{33})$\\
(63) & (0001101011) & 8 &$a_1=a_2=a_3=b_{22}=b_{33}=n_{31}=n_{31}'=g_{31}'=0$ & $b_{21}b_{23}$\\
\end{longtable}
\vspace{-3ex}\noindent
  $(^1)$ $a_1a_2d_{13}(d_{13}k_{12}-a_2b_{13}b_{22})\neq 0$\\
  $(^2)$ $b_{11}b_{33}(a_1(2b_{21}b_{33}^2-5b_{13}b_{21}b_{33}+
  2b_{13}^2b_{21}+b_{11}b_{23}b_{33})-2a_2b_{11}d_{13}^2)\neq 0$\\
  $(^3)$ $d_{13}d_{31}j_{23}({{ b_{11}}}^{3}(2\,{ b_{32}}\,{ b_{13}}+{ b_{22}}\,{ b_{33}}-3\,{ b_{33}}\,{
  b_{32}} )-{ b_{22}}\,{{ b_{31}}}^{3}\, d_{13}- {{ b_{11}}}^{2}{ b_{21}}({ b_{32}}\,{ b_{13}}+2\,{ b_{22}}\,
  { b_{33}}-3\,{ b_{33}}\,{ b_{32}} )+{b_{13}}\,{{ b_{21}}}^{2}\,{ b_{31}}\, d_{32}+ 
  { b_{21}}{{ b_{31}}}^{2}\,d_{13}\,d_{32}'' + {{ b_{11}}}^{2}{ b_{31}}(3\,{ b_{22}}\,{ b_{33}}+
  2\,{ b_{33}}\,{ b_{32}}-4\,{ b_{22}}\,{ b_{13}}-{ b_{32}}\,{ b_{13}} )-
  { b_{11}}{ b_{33}}{{ b_{21}}}^{2}\, d_{32}+4\,{ b_{11}}\, { b_{22}}{{ b_{31}}}^{2}\,d_{13}+
  2\,{ b_{11}}\,{b_{21}}\,{ b_{31}}(2\,{b_{13}}\,{ b_{22}}-{ b_{22}}\,{ b_{33}}-{ b_{13}}\,{ b_{32}} ))\neq 0$\\
  $(^4)$ $d_{13}j_{13}(4\,{ b_{11}}\,{ b_{23}}\, d_{32}^{2}-{{ 
  b_{12}}}^{2}{ b_{23}}\,d_{31}-{ b_{12}}\,{ b_{23}}{ b_{31}}\,(3\,{ b_{32}}-4\,{ b_{22}} )+{ b_{11}}\,{ b_{12}}\,{ b_{23}}
  \, (3\,{ b_{32}}-4\,{ b_{22}} )+{b_{12}}\,{ b_{13}}\,{ b_{22}}\,{ b_{31}}- { b_{11}}\,{ b_{13}}(2\,{{ b_{32}}}^{2}-{ 
  b_{32}}\,{ b_{22}}-2\,{{ b_{22}}}^{2} )-{b_{11}}\,{ b_{12}}\,{ b_{13}}\,{ b_{22}}- { b_{13}}\,{ b_{31}}d_{32}''(2\,{ b_{32}}-3\,{ b_{22}}) )\neq 0$\\
  $(^5)$ $b_{21}b_{33}d_{23}d_{13}''d_{31}''d_{23}''d_{32}''g_{21}\neq 0$\\
  $(^6)$   $b_{22}b_{33}d_{23}''(b_{12}d_{23}'d_{23}''+b_{22}b_{13}b_{33})\neq 0$\\
  $(^7)$ $a_1a_2a_3d_{23}''(b_{12}d_{23}'d_{23}''+b_{22}(b_{13}b_{33}-2d_{23}^2))\neq 0$\\
  $(^8)$ $b_{33}d_{31}''(b_{33}d_{21}''-b_{23}d_{31}'')\neq 0$\\
  $(^9)$  $b_{22}b_{33}d_{12}''d_{13}''d_{21}''(2b_{12}-3b_{22})\neq 0$\\
  $(^{10})$  $a_1a_2a_3b_{33}(3b_{12}b_{33}+b_{32}d_{13}'')\neq 0$\\
  $(^{11})$  $b_{23}b_{32}(b_{13}b_{32}+3b_{12}b_{33})\neq 0$\\
  $(^{12})$ $b_{11}b_{33}d_{13}''(b_{33}b_{32}^2+(4b_{13}-5b_{33})b_{22}^2)\neq 0$\\
  $(^{13})$ $b_{21}b_{23}b_{33}(b_{23}^2-b_{33}d_{23})\neq 0$
}

\bigskip
\noindent{\bf Theorem 5.} {\it Let $f_1=x$, $f_2=y$ and $f_3=z$. If a LV3 has a
 fourth algebraic solution, $f_4=0$ of degree 2 not passing through the origin, then 
a first integral $|f_1|^{\la_1}|f_2|^{\la_2}|f_3|^{\la_3}|f_4|^{\la_4}$  can be obtained 
when the conditions given in Table 3 are fulfilled.}

{\small
\begin{longtable}{|c|c|l|l|l|}
  \caption{First integral conditions of Theorem 5.} \\
  \hline  
  st. & $\overrightarrow{\nu}$ & n & conditions =0&  conditions  $\neq  0$\\
  \hline
  \endfirsthead
  \caption[]{First integral conditions of Theorem 5 (continued).} \\
  \hline  
  st. & $\overrightarrow{\nu}$ & n & conditions =0&  conditions  $\neq  0$\\
  \hline
  \endhead
  \hline
  \endfoot
$(1)$ & $(1011000011)$ &  4  & $r_{23}' =b_{31}=b_{21}=b_{32}=0$ &$a_2a_3c_{23}''$\\
$(2)$ & $(1110110100)$ &  5  & $b_{13}=b_{23}=b_{33}=r_{12}'=r_{21}'$ =0&$(^{*})$\\
$(3)$ & $(1110100100)$ &  5  & $c_{12}''=b_{13}=b_{23}=d_{12}=d_{21}=0$ &$|a_1|+|b_{11}|+|b_{22}|$\\
$(4)$ & $(1001000111)$ &  5  & $a_2=b_{21}=b_{22}=b_{31}=d_{23}''=0$ &$a_3$\\
$(5)$ & $(1111111111)$ &  6  & $k_{12}=r_{12}=r_{23}'=r_{31}'=r_{32}'=r_{13}=0$ &$a_1a_2c_{23}'''c_{32}'''$\\
$(6)$ & $(1111111111)$ &  6  & $k_{23}=k_{32}=r_{12}'=r_{31}'=r_{21}'=r_{13}'=0$ &$a_1a_2a_3c_{23}b_{11}b_{22}b_{33}$\\
$(7)$ & $(1111111111)$ &  6  & $k_{23}=k_{32}=r_{12}'=r_{31}=r_{21}'=r_{13}'=0$ &$a_1a_2a_3b_{13}b_{22}$\\
$(8)$ & $(1111111111)$ &  6  & $k_{31}=k_{13}=k_{32}'=r_{12}'=r_{23}'=r_{21}'=0$ &$a_1a_2a_3$\\
$(9)$ & $(1011011111)$ &  6  & $e_{21}=e_{31}=r_{12}'=r_{23}=k_{32}=r_{13}'=0$ &$a_2a_3c_{23}c_{12}''c_{13}''$\\
$(10)$ & $(1111000111)$ &  6  & $d_{31}'=d_{21}'=r_{12}'=r_{23}'=r_{32}'=r_{13}'=0$ &$a_1a_2a_3v_{231}'$\\
$(11)$ & $(1001001011)$ &  6  & $e_{31}=e_{32}=g_{31}''=r_{12}=r_{13}'=r_{23}'=0$ &$a_3c_{12}c_{13}''c_{23}''$\\
$(12)$ & $(1111001111)$ &  6  & $b_{31}=d_{21}'=k_{23}=k_{32}=r_{12}'=r_{13}'=0$ &$a_1a_2a_3c_{13}''$\\
$(13)$ & $(1011001111)$ &  6  & $b_{21}=e_{31}=r_{23}'=n_{23}'=r_{32}'=r_{13}'$ =0&$a_2a_3c_{13}''c_{13}'''v_{132}'$\\
$(14)$ & $(1111000011)$ &  6  & $b_{32}=d_{31}'=d_{21}'=r_{12}=r_{23}=r_{13}'=0$ &$a_1a_2a_3c_{12}'$\\
$(15)$ & $(1011001001)$ &  6  & $b_{21}=d_{12}'=r_{23}' =e_{31}=d_{32}'=r_{13}'=0$ &$a_2a_3c_{13}''v_{312}'$\\
(16)   & $(1011001111)$ &  6  &$b_{21}=e_{31}=r_{23}=r_{12}'=r_{13}'=k_{23}=0$ & $a_2a_3c_{13}''$\\
(17)   & $(1111001001)$ &  6  &$b_{31}=r_{12}=r_{23}'=r_{13}=d_{12}'=g_{31}=0$ & $a_1a_2a_3c_{21}'c_{21}'''$\\
$(18)$ & $(1011010110)$ &  6  & $b_{23} =b_{31}=e_{21}=r_{12}'=r_{32}=s_{13}'=0$ &$a_2a_3c_{12}''w_{312}''$\\
$(19)$ & $(1011001011)$ &  6  & $b_{21}=b_{32}=e_{31}=o_{123}=r_{23}=r_{13}'=0$ &$a_2a_3b_{13}c_{13}''w_{213}''$\\
$(20)$ & $(1111001111)$ &  6  & $b_{31}=b_{32}=r_{31}=r_{32}=r_{12}'=d_{21}'=0$ &$a_1a_2a_3b_{11}b_{22}b_{33}$\\
$(21)$ & $(1011001010)$ &  6  & $b_{13}=b_{21}=d_{23}'=e_{31}=m_{12}=r_{32}'=0$ &$a_1a_2a_3c_{13}''v_{132}''$\\
$(22)$ & $(1011011000)$ &  6  & $b_{12}=b_{13}=d_{23}=d_{32}=e_{21}=e_{31}=0$ &$a_2a_3c_{23}c_{12}''c_{13}''$\\
(23)   & $(1111001011)$ &  6  & $b_{31}=b_{32}=d_{12}=d_{21}=r_{13}=r_{23}'=0$ & $a_1a_3a_2c_{23}''$\\
(24)   & $(1111001011)$ &  6  & $b_{31}=b_{32}=r_{12}=r_{23}=r_{13}=r_{12}'''=0$ & $a_1a_3a_2b_{11}b_{22}c_{12}$\\
(25)   & $(1011010110)$ &  6  & $b_{31}=b_{23}=r_{12}'=e_{21}=r_{32}''=s_{13}'=0$ & $a_2a_3c_{12}c_{23}'''v_{123}''$\\
$(26)$ & $(1011001010)$ &  6  & $b_{23}=b_{21}=b_{32}=b_{13}=e_{31}=s_{12}=0$ &$a_2a_3c_{13}''$\\
$(27)$ & $(1011011111)$ &  6  & $c_{32}=r_{12}'=r_{23}=e_{31}=e_{21}=r_{13}'=0$ &$a_2a_3c_{12}''c_{13}''c_{12}'''c_{13}'''d_{32}$\\
$(28)$ & $(1001010011)$ &  6  & $c_{12}''=b_{31}=r_{23}'=e_{12}=e_{21}=e_{32}=0$ &$a_3c_{13}c_{13}'g_{12}''k_{13}$\\
$(29)$ & $(1001011001)$ &  6  & $c_{12}''=b_{32}=e_{12}=e_{21}=e_{31}=r_{13}'=0$ &$a_3c_{13}''g_{12}''m_{23}$\\
$(30)$ & $(1001010010)$ &  6  & $c_{12}''=b_{23}=b_{31}=e_{12}=e_{21}=e_{32}=0$ &$a_3b_{13}d_{13}c_{13}$\\
$(31)$ & $(1011001111)$ &  6  & $a_2=b_{22}=b_{21}=e_{31}=n_{32}'=r_{13}'=0$ &$a_3b_{13}b_{23}c_{13}''$\\
(32) & $(1111000111)$   & 7 &$c_{23}'''=d_{21}'=d_{31}'=e_{32}'=e_{23}''=r_{12}'=r_{13}'=0$ & $a_1a_2a_3c_{12}'''$\\
(33) & $(1111000111)$   & 7 &$c_{13}'''=b_{13}=d_{21}'=d_{31}'=k_{23}=k_{32}=r_{12}'=0$ & $a_1a_2a_3c_{12}'''$\\
(34) & $(1111011111)$   & 7 &$c_{21}'''=b_{21}=b_{31}=r_{12}'=r_{31}=r_{23}'=n_{23}=0$ & $a_1a_2a_3$\\
(35) & $(1111001001)$   & 7 &$c_{21}'''=b_{21}=b_{31}=r_{13}=d_{12}'=g_{31}=r_{23}'=0$ & $a_1a_2a_3$\\
(36) & $(1111000111)$ & 7 &$c_{23}''=d_{21}'=e_{23}=d_{31}'=e_{32}=r_{12}'=r_{13}'=0$ & $a_1a_2a_3$\\
(37) & $(1111001100)$ & 7 &$w_{231}=b_{13}=b_{31}=d_{21}'=d_{23}'=r_{12}'=r_{32}'=0$ & $a_1a_3c_{31}''$\\
(38) & $(1011010110)$ & 7 &$w_{312}=b_{13}=b_{23}=b_{31}=e_{21}=n_{23}'=r_{23}=0$ & $a_2c_{12}''$\\
$(39)$ & $(1111111111)$ &  7  & $v_{123}=r_{12}'=r_{23}'=r_{31}'=r_{21}'=r_{32}'=r_{13}'=0$ &$a_1a_2a_3b_{11}b_{22}b_{33}$\\
$(40)$ & $(1111011010)$ &  7  & $v_{123}=b_{12}=b_{23}=b_{31}=b_{21}=b_{32}=b_{13}=0$ &$a_1a_2c_{12}''$\\
$(41)$ & $(1001011011)$ &  7  & $c_{12}''=e_{31}=e_{32}=g_{13}=g_{23}=r_{13}'=r_{23}'=0$ &$c_{13}c_{13}''c_{13}''$\\
$(42)$ & $(1111011111)$ &  7  & $c_{31}'''=b_{31}=b_{21}=r_{12}=r_{23}'=r_{32}'=r_{13}'=0$ &$a_1a_2a_3$\\
$(43)$ & $(1111001110)$ &  7  & $c_{32}'''=b_{23}=b_{31}=b_{32}=b_{13}=r_{12}'=d_{21}'=0$ &$a_1a_2a_3c_{13}$\\
$(44)$ & $(1011011011)$ &  7  & $c_{23}=b_{12}=b_{32}=e_{31}=e_{21}=r_{13}'=r_{23}'=0$ &$a_1a_2a_3c_{12}''$\\
$(45)$ & $(1011011111)$ &  7  & $w_{123}=r_{12}'=r_{23}'=e_{31}=e_{21}=r_{32}'=r_{13}'=0$ &$a_2a_3c_{23}c_{12}''c_{13}''$\\
$(46)$ & $(1111011111)$ &  7  & $w_{123}=b_{31}=b_{21}=r_{12}=r_{23}'=r_{32}'=r_{13}=0$ &$a_1a_2a_3b_{23}b_{32}c_{32}''$\\
$(47)$ & $(1011011010)$ &  7  & $w_{123}=b_{12} =b_{23}=b_{32}=b_{13}=e_{31}=g_{23}=0$ &$a_2a_3c_{12}''c_{31}''c_{31}'''$\\
$(48)$ & $(1111010111)$ &  7  & $w_{231}'''=b_{21}=d_{31}'=r_{12}=r_{23}'=r_{32}'=r_{13}'=0$ &$a_1a_2a_3$\\\
$(49)$ & $(1111001110)$ &  7  & $w_{123}'''=b_{23}=b_{31}=b_{13}=d_{21}'=r_{12}'=r_{32}=0$ &$a_1a_2a_3c_{12}'''$\\
$(50)$ & $(1011011011)$ &  7  & $w_{213}'=b_{12} =b_{32}=e_{31}=e_{21}=r_{23}=r_{13}'=0$ &$a_3c_{13}c_{13}''$\\
(51)   & $(1011111111)$ & 7 &$a_1=b_{11}=n_{12}=n_{31}=n_{23}'=n_{31}'=k_{23}=0$ & $a_2a_3$\\
(52)   & $(1011111111)$ & 7 &$a_1=b_{11}=d_{12}''=n_{31}=n_{23}=l_{23}'=r_{23}'=0$ & $a_2a_3$\\
(53)   & $(1111000111)$ & 7 &$a_3=b_{33}=d_{21}'=d_{31}'=d_{32}''=l_{12}=r_{12}'=0$ & $a_1a_2c_{12}$\\
$(54)$ & $(1111000111)$ &  7  & $a_2=b_{22}=d_{21}'=d_{31}'=l_{13}=r_{13}'=r_{23}'=0$ &$a_1a_3c_{13}c_{13}'''$\\
$(55)$ & $(1011001001)$ &  7  & $a_3=b_{21}=b_{33}=d_{12}'=e_{31}=g_{31}=l_{21}=0$ &$a_1a_2c_{12}$\\
$(56)$ & $(1111111111)$ &  7  & $a_1=b_{11}=b_{13}=r_{12}'=r_{23}'=r_{32}=l_{23}$ =0&$a_2a_3$\\
$(57)$ & $(1111111111)$ &  7  & $a_1=b_{11}=b_{13}=r_{12}'=r_{23}'=r_{32}'=l_{23}$ =0&$a_2a_3$\\
$(58)$ & $(1111111111)$ &  7  & $a_1=b_{11}=b_{12}=k_{23}'=r_{32}=r_{13}'=l_{23}$ =0&$a_2a_3$\\
(59) & (1011011111) & 7 &$a_2=b_{22}=b_{23}=e_{31}=g_{23}=l_{13}=n_{23}'=0$ & $a_1a_3c_{13}''$\\
(60) & (1011010111) & 7 &$a_3=b_{31}=b_{33}=e_{21}=d_{32}''=r_{12}'=r_{12}^{iv}=0$ & $a_2c_{12}''c_{12}'''$\\
(61) & (1011001111) & 7 &$a_3=b_{21}=b_{33}=e_{31}=d_{32}''=l_{12}=k_{12}'=0$ & $a_1a_2c_{12}$\\
(62) & (1111000111) & 7 &$a_2=b_{22}=b_{23}=l_{13}=d_{21}'=d_{31}'=r_{13}'=0$ & $a_1a_3$\\
(63) & (1111000111) & 7 &$a_2=b_{22}=b_{32}=d_{21}'=d_{31}'=r_{13}'=o_{123}=0$ & $a_1a_3$\\
(64) & (1111000110) & 7 &$a_2=b_{22}=b_{23}=d_{21}'=g_{32}=l_{31}=r_{31}=0$ & $a_1a_3c_{13}'$\\
(65) & (1111001001) & 7 &$a_1=b_{11}=b_{31}=d_{12}'=d_{13}=g_{31}=r_{23}'=0$ & $a_2a_3$\\
(66) & (1111000011) & 7 &$a_3=b_{32}=b_{33}=d_{21}'=d_{31}'=l_{12}=r_{12}=0$ & $a_1a_2c_{12}'$\\
(67) & (1111001111) & 7 &$a_2=b_{22}=b_{31}=b_{32}=d_{21}'=d_{23}=r_{13}=0$ & $a_1a_3c_{13}''$\\
(68) & (1111001111) & 7 &$a_2=b_{22}=b_{23}=b_{31}=d_{21}'=n_{23}'=r_{13}'=0$ & $a_1a_3c_{13}''$\\
(69) & (1111010111) & 7 &$a_3=b_{21}=b_{32}=b_{33}=d_{31}'=l_{12}=r_{12}'=0$ & $a_1a_2c_{12}''$\\
(70) & (1011010111) & 7 &$a_3=b_{13}=b_{31}=b_{33}=e_{21}=r_{12}'=n_{23}'=0$ & $a_1a_2c_{12}''$\\
(71) & (1011001111) & 7 &$a_3=b_{33}=b_{21}=b_{32}=e_{31}=n_{23}'=l_{12}=0$ & $a_1a_2$\\
(72) & (1011010110) & 7 &$a_3=b_{23}=b_{31}=b_{33}=e_{21}=d_{32}=r_{12}'=0$ & $a_2c_{12}''$\\
(73) & (1011001011) & 7 &$a_3=b_{21}=b_{32}=b_{33}=e_{31}=l_{12}=o_{123}=0$ & $a_1a_2c_{12}'$\\
(74) & (1011010111) & 7 &$a_2=a_3=b_{22}=b_{31}=b_{33}=e_{21}=n_{23}'=0$ & $b_{12}g_{12}''$\\
(75) & (1010011100) & 8 &$c_{13}''=b_{13}=b_{23}=b_{33}=e_{21}=e_{31}=k_{32}=r_{12}'=0$ & $a_2c_{32}$\\
(76) & (1001011001) & 8 &$c_{12}''=b_{12}=b_{22}=b_{32}=e_{31}=g_{32}=k_{23}=r_{13}'=0$ & $a_3c_{13}''$\\
(77) & (1001011011) & 8 &$c_{12}''=c_{13}'''=b_{13}=e_{31}=e_{32}=g_{31}=g_{32}=r_{23}'=0$ & $a_1a_2a_3$\\
(78) & (1111010111) & 8 &$c_{12}'''=c_{32}''=b_{21}=b_{23}=b_{32}=d_{31}'=e_{12}=r_{13}'=0$ & $a_1a_2a_3$\\
(79) & (1111010111) & 8 &$c_{12}''=c_{13}=b_{12}=b_{21}=b_{23}=b_{32}=d_{31}'=r_{13}'=0$ & $a_1a_2a_3$\\
(80) & (1011100110) & 8 &$a_1=b_{11}=b_{21}=b_{23}=b_{31}=d_{12}=d_{13}'=r_{32}'=0$ & $a_2c_{23}''b_{22}b_{33}$\\
(81) & (1011011111) & 8 &$a_2=c_{31}'''=b_{22}=d_{23}''=e_{31}=g_{23}=r_{13}'=o_{123}=0$ & $a_1a_3$\\
(82) & (1011011111) & 8 &$a_3=c_{21}'''=b_{33}=e_{31}=g_{23}=n_{23}=r_{12}'=o_{123}=0$ & $a_1a_2$\\
(83) & (1111001111) & 8 &$a_2=c_{31}'''=b_{22}=b_{31}=d_{21}'=d_{13}'=g_{31}'''=d_{23}''=0$ & $a_1a_3$\\
(84) & (1111010111) & 8 &$a_3=c_{21}'''=b_{21}=b_{33}=d_{31}'=d_{32}''=g_{21}'''=r_{12}=0$ & $a_1a_2$\\
(85) & (1011011011) & 8 &$a_3=c_{12}'''=b_{12}=b_{32}=b_{33}=e_{31}=g_{23}=g_{12}'''=0$ & $a_1a_2$\\
(86) & (1111000011) & 8 &$a_3=c_{12}'''=b_{12}=b_{32}=b_{33}=d_{21}'=d_{31}'=g_{12}'''=0$ & $a_1a_2$\\
\end{longtable}
\vspace{-3ex}
\noindent$(^{*})\,\,a_1a_2c_{12}'''c_{21}'''(a_3{c_{12}''}^2b_{12}b_{21}-
              a_1^2c_{12}'''b_{21}b_{32}-a_2^2c_{21}'''b_{12}b_{31})\neq 0$
}

\section{Integrability}

We say that a 3-dimensional system like (\ref{l1}) is {\it integrable} if it has
two independent first integrals. Theorems 6--10 concern integrable LV3
having at least one first integral formed with linear algebraic solutions and
Theorem 11 is for integrable cases with algebraic solutions being both quadratic.
In these theorems we indicate the nature of the second first integral computed using 
the notation P1(st) and Tk(st), respectively for Proposition 1, statement st and Theorem k, 
statement st. and Q is for quadratic first integrals which do not belong to the statements of 
Theorems 4 and 5. 

\newpage

\noindent{\bf Theorem 6.} {\it An LV3 is integrable with two first integrals
one of which being of the type of Theorem 2(1) in the cases of Table 4.}
{\small
\begin{longtable}{|c|c|l|l|l|}
  \caption{Integrability conditions of Theorem 6.}\\
  \hline  
   st. &  n & conditions =0&  conditions $\neq  0$ & f.i. type\\
  \hline
  \endfirsthead
  \caption[]{Integrability conditions of Theorem 6 (continued).}\\
  \hline  
   st. &  n & conditions =0&  conditions $\neq  0$ & f.i. type\\
  \hline
  \endhead
  \hline
  \endfoot
(1) &  4  &  $r_{12}=r_{23}=r_{31}=a_1b_{22}d_{23}+a_2h_{32}=0 $ & $a_1 a_2 a_3b_{11}b_{22}b_{33}k_{12}$ & P1(5)\\
(2) &  5  &  $b_{13}=b_{23}=b_{33}=r_{12}=a_3b_{22}d_{21}+a_2h_{12}=0 $ & $a_1 a_2 b_{11}b_{22}$ & T2(4)\\
(3) &  5  &  $a_3=b_{33}=r_{12}=l_{12}=o_{123}=0 $ & $a_1 a_2 b_{11}b_{22}$ &T2(2)\\
(4) &  6  &  $c_{12}=c_{23}=n_{12}=n_{23}=n_{31}=o_{123}=0 $ & $b_{12}d_{21}d_{31}''$ & Q\\
(5) &  6  &  $b_{21}=b_{31}=k_{23}=k_{32}=r_{12}'=r_{13}'$ & $a_1a_2c_{12}''c_{13}''$ & Q\\
(6) &  6  &  $b_{31}=b_{32}=d_{12}=d_{21}=r_{23}'=r_{13}'$ & $a_1 a_2$ & Q\\
(7) &  6  &  $a_1=c_{23}=b_{11}=n_{23}=n_{23}'=g_{32}=0 $ & $b_{13}d_{13}d_{23}d_{23}''j_{12}$ & Q\\
(8) &  6  &  $a_3=b_{33}=b_{31}=b_{32}=r_{12}=o_{123}=0 $ & $a_1 a_2 k_{12}$ & P1(7)\\
(9) &  7  &  $a_2=b_{22}=b_{21}=b_{23}=b_{31}=l_{13}=n_{23}'=0 $ & $a_1 a_3 c_{13}''$ & Q\\
(10) &  7  &  $c_{23}=b_{12}=b_{22}=b_{32}=d_{23}=g_{23}=r_{31}=0 $ & $a_1 b_{33} k_{13}$ & T2(4) \\
(11) &  7  &  $a_2=a_3=b_{22}=b_{23}=b_{32}=b_{33}=g_{23}=0 $ & $b_{11}b_{21}$ & P1(3,7)\\
(12) &  7  &  $a_3=a_1=b_{33}=b_{31}=b_{12}=b_{13}=b_{11}=0 $ & $b_{32}d_{32}$ & P1(7)\\
(13) &  7  &  $a_1=a_2=b_{11}=b_{12}=b_{21}=b_{22}=d_{13}''=0 $ & $a_3 b_{33} d_{23}$ & Q\\
(14) &  7  &  $a_1=a_2=b_{11}=b_{12}=b_{21}=b_{22}=j_{12}=0 $ & $a_3 d_{23} d_{23}''$ & Q\\
(15) &  7  &  $a_2=a_3=b_{23}=b_{22}=b_{33}=d_{32}''=i_{23}'''=0 $ & $b_{11} b_{13} d_{31}$ & Q\\
(16) &  7  &  $a_1=a_3=b_{11}=b_{13}=b_{31}=b_{33}=i_{13}'''=0 $ & $a_2 b_{32} g_{13}$ & Q\\
(17) &  7  &  $a_2=a_3=b_{22}=b_{23}=b_{32}=b_{33}=i_{32}=0 $ & $b_{11} b_{12} b_{21} d_{21}$ & Q\\
(18) &  7  &  $a_1=a_2=a_3=b_{33}=e_{31}=e_{32}=o_{123}=0 $ & $b_{11}b_{22}b_{23}b_{32}$ & T2(2)\\
(19) &  7  &  $a_1=a_2=a_3=b_{13}=d_{31}''=i_{31}=o_{123}=0 $ & $b_{11}b_{33}d_{12}d_{21}$ & Q\\
(20) &  7  &  $a_1=a_2=a_3=b_{13}=b_{31}=i_{31}'=o_{123}=0 $ & $b_{11}b_{32}$ & Q\\
(21) &  7  &  $a_1=a_2=a_3=b_{31}=b_{32}=b_{33}=o_{123}=0 $ & $b_{11} d_{21} (b_{11} b_{22}-b_{12} b_{21})$ & P1(4)\\
(22) &  7  &  $a_1=a_2=a_3=b_{11}=b_{22}=b_{33}=o_{123}=0 $ & $$ & T2(2)\\
(23) &  7  &  $a_1=a_2=a_3=b_{12}=b_{13}=b_{22}=b_{32}=0 $ & $b_{11} d_{31}$ & P1(3)\\
(24) &  8  &  $c_{23}=g_{23}=d_{12}=d_{21}=d_{32}=b_{13}=b_{23}=b_{33}=0 $ & $a_1 a_2 a_3 c_{12}$ & T2(5)\\
(25) &  8  &  $c_{12}=c_{32}'=d_{12}=d_{21}=d_{13}=d_{31}=d_{23}=d_{32}=0 $ & $b_{13}$ & T4(15)\\
(26) &  8  &  $c_{31}'=b_{12}=b_{13}=b_{22}=b_{23}=b_{32}=b_{33}=d_{31}=0$ & $l_{32}$ & T2(5)\\
(27) &  8  &  $c_{23}=b_{12}=b_{13}=b_{22}=b_{23}=b_{32}=b_{33}=g_{23}=0 $ & $a_1 b_{11} k_{21}$ & T2(7)\\
(28) &  8  &  $c_{31}''=b_{11}=b_{13}=b_{21}=b_{23}=b_{31}=r_{23}=r_{12}''=0 $ & $a_2$ & T2(4)\\
(29) &  8  &  $a_1=c_{23}=b_{11}=b_{12}=b_{22}=b_{32}=d_{23}=g_{23}=0 $ & $b_{33}$ & T2(6)\\
(30) &  8  &  $a_3=c_{12}'''=b_{12}=b_{21}=b_{31}=b_{32}=b_{33}=g_{12}'''=0 $ & $a_1 a_2$ & Q\\
(31) &  8  &  $a_2=c_{31}''=b_{13}=b_{22}=b_{31}=d_{21}'=d_{23}'=g_{31}''=0 $ & $a_1 a_3$ & Q\\
(32) &  8  &  $a_1=b_{11}=b_{21}=b_{23}=b_{31}=d_{12}''=d_{13}=r_{32}'=0 $ & $a_2 a_3 c_{23}''$ & Q\\
(33) &  8  &  $a_1=c_{23}=b_{11}=b_{21}=b_{31}=d_{13}=j_{13}=n_{32}=0 $ & $d_{23}'' b_{22} b_{33}$ & T4(2)\\
(34) &  8  &  $a_2=c_{13}'''=b_{13}=b_{21}=b_{22}=b_{23}=b_{31}=g_{13}'''=0 $ & $a_1a_3$ & Q\\
(35) &  8  &  $a_2=c_{13}''=b_{11}=b_{21}=b_{22}=b_{31}=d_{13}=g_{31}''=0 $ & $b_{23}g_{12}$ & P1(7)\\
(36) &  8  &  $a_3=c_{21}''=b_{12}=b_{21}=b_{31}=b_{32}=b_{33}=g_{21}''=0 $ & (*) & P1(7)\\
(37) &  8  &  $a_1=a_3=b_{11}=b_{13}=b_{21}=b_{31}=b_{33}=d_{12}=0 $ & $b_{22} b_{32}$ & T2(6)\\
(38) &  8  &  $a_1=a_3=b_{11}=b_{12}=b_{13}=b_{31}=b_{33}=d_{32}'=0 $ & $a_2$ & Q\\
(39) &  8  &  $a_2=a_3=b_{12}=b_{22}=b_{23}=b_{32}=b_{33}=g_{23}=0 $ & $b_{21}$ & P1(8)\\
(40) &  8  &  $a_2=a_1=b_{22}=b_{21}=b_{12}=b_{11}=d_{23}'=g_{12}=0 $ & (**) & P1(10)\\
(41) &  8  &  $a_1=a_2=b_{11}=b_{12}=b_{21}=b_{22}=b_{31}=d_{13}=0 $ & $a_3 g_{12} b_{23} b_{13} b_{33}$ & P1(8)\\
(42) &  8  &  $a_1=a_2=a_3=b_{33}=n_{31}=n_{23}=n_{23}'=o_{123}=0 $ & $b_{12} b_{13} b_{23}$ & T2(2)\\
(43) &  8  &  $a_1=a_2=a_3=b_{12}=b_{22}=b_{31}=b_{32}=b_{33}=0 $ & $b_{11}$ & T(2)\\
(44) &  8  &  $a_1=a_2=a_3=b_{11}=b_{21}=b_{22}=b_{23}=b_{31}=0 $ & $b_{33}$ & P1(4)\\
(45) &  8  &  $a_1=a_2=a_3=b_{21}=b_{33}=d_{31}''=n_{23}=o_{123}=0 $ & $b_{23}$ & P1(4)\\
(46) &  8  &  $a_1=a_2=a_3=b_{33}=d_{32}''=g_{12}''=n_{31}=o_{123}=0 $ & $b_{11}b_{23}$ & P1(4)\\
(47) &  8  &  $a_1=a_2=a_3=b_{11}=b_{13}=b_{31}=b_{33}=d_{32}''=0 $ & $b_{22}$ & T2(2)\\
(48) &  8  &  $a_1=a_2=a_3=b_{11}=b_{13}=b_{31}=b_{32}=b_{33}=0 $ & $b_{12}$ & P1(4)\\
(49) &  8  &  $a_1=a_2=a_3=b_{21}=b_{22}=b_{23}=d_{31}=d_{13}=0 $ & $g_{31}$ & Q\\
(50) &  8  &  $a_1=a_2=a_3=b_{22}=b_{33}=e_{31}=g_{23}=n_{23}'=0 $ & $b_{12} b_{23} b_{32} g_{31}'''$ & Q\\
(51) &  8  &  $a_1=a_2=a_3=b_{11}=b_{33}=d_{32}''=g_{23}=n_{23}'=0 $ & $b_{13} b_{23} b_{22} g_{12}''$ & T4(1)\\
(52) &  8  &  $a_1=a_2=a_3=b_{13}=b_{23}=d_{31}''=d_{32}''=n_{12}'=0 $ & $b_{11} b_{21} b_{33}$ & Q\\
(53) &  8  &  $a_1=a_2=a_3=b_{22}=b_{33}=d_{31}''=g_{13}=n_{31}'=0 $ & $b_{21} b_{23} d_{21}$ & Q\\
(54) &  9  &  $c_{12}=c_{31}'=b_{32}=b_{12}=b_{22}=d_{21}=d_{13}=d_{31}=d_{23}=0 $ & $|a_1|+|b_{11}|+|b_{33}|$ & T2(5)\\
(55) &  9  &  $c_{12}=c_{13}=b_{11}=b_{21}=b_{31}=b_{32}=d_{13}''=d_{12}''=n_{23}=0 $ & $b_{33} b_{22}$ & Q\\
(56) &  9  &  $c_{13}''=c_{23}''=b_{11}=b_{21}=b_{31}=d_{12}=d_{23}=n_{32}=d_{13}=0 $ & $a_1a_2a_3$ & Q\\
(57) &  9  &  $a_1=c_{23}''=b_{11}=b_{12}=b_{13}=b_{23}=b_{33}=e_{32}=g_{23}''=0 $ & $$ & T2(1)\\
(58) &  9  &  $a_3=c_{12}''=b_{12}=b_{13}=b_{21}=b_{23}=b_{31}=b_{32}=b_{33}=0 $ & $|a_1|+|b_{11}|+|b_{22}|$ & T2(4)\\
(59) &  9  &  $a_3=c_{21}''=b_{23}=b_{33}=b_{13}=d_{32}=d_{31}=d_{12}=d_{21}=0 $ & $a_1a_2$ & Q\\
(60) &  9  &  $a_1=c_{23}'''=b_{11}=b_{21}=b_{23}=b_{31}=b_{32}=d_{12}=d_{13}''=0 $ & $a_2a_3$ & Q\\
(61) &  9  &  $a_2=a_1=b_{22}=b_{21}=b_{32}=b_{12}=b_{13}=b_{11}=d_{23}=0 $ & $a_3$ & P1(8)\\
(62) &  9  &  $a_1=a_3=b_{11}=b_{13}=b_{23}=b_{31}=b_{33}=d_{12}''=d_{32}=0 $ & $a_2$ & Q\\
(63) &  9  &  $a_1=a_2=a_3=b_{11}=b_{22}=b_{33}=b_{23}=b_{32}=g_{23}=0 $ & $|b_{21}|+|b_{12}|+|b_{13}|$ & T2(2) \\
\end{longtable}
\vspace{-3ex}
\noindent (*) $|a_1|+|b_{11}|+|b_{22}|+|b_{23}|\neq 0\qquad$ (**) $|a_3|+|b_{31}|+|b_{32}|+|b_{33}|\neq 0$
}

\bigskip
\noindent{\bf Theorem 7.} {\it An LV3 is integrable with two first integrals
one of which being of the type of Theorem 2(2) in the cases of Table 5, which 
exclude those already mentioned in Theorem 6.}
\begin{table}[htbp]
\vspace{-1em}
\caption{Integrability conditions of Theorem 7.}
\vspace{1em}
\small
\centering
\begin{tabular}{|c|c|l|l|l|}
\hline  
 st. &  n & conditions =0&  conditions $\neq  0$ & f.i. type\\
\hline
(1) &  5  &  $c_{12}=c_{13}=d_{12}=j_{12}=o_{123}=0$ &  & Q\\
(2) &  6  &  $c_{12}=c_{23}=b_{12}=b_{32}=r_{31}=o_{123}=0$ & $b_{11}b_{31}b_{33}d_{23}$ & Q\\
(3) &  6  &  $c_{12}=c_{13}=o_{123}=d_{31}'=d_{21}'=r_{32}=0$ & $b_{11}b_{33}$ & T4(49)\\
(4) &  6  &  $c_{12}=c_{13}=b_{33}=n_{12}'=n_{23}'=o_{123}=0$ & $d_{31}$ & T2(4)\\
(5) &  6  &  $a_1=a_2=a_3=g_{23}=h_{23}=o_{123}=0 $ & $b_{22} b_{33} d_{13} d_{32}$ & T2(3)\\
(6) &  7  &  $c_{12}=c_{13}=b_{13}=b_{23}=b_{31}=b_{32}=r_{12}=0$ & $a_1a_2a_3b_{22}b_{12}b_{21}$ & Q\\
(7) &  7  &  $a_1=a_2=a_3=b_{13}=b_{23}=j_{23}=g_{31}=0 $ & $(^{*})$ & T2(3)\\
(8) &  7  &  $a_1=a_2=a_3=b_{12}=b_{13}=b_{23}=o_{123}=0 $ & $b_{33} d_{21} j_{23}$ & T2(3)\\
(9) &  7  &  $a_1=a_2=a_3=b_{22}=b_{33}=g_{23}=o_{123}=0 $ & $b_{11}b_{23}$ & T2(3)\\
(10) &  8  &  $c_{12}=c_{13}=b_{11}=b_{22}=b_{31}=d_{23}''=g_{21}'=g_{31}''=0$ & $b_{21}b_{33}$ & T2(4)\\
(11) &  8  &  $a_1=a_2=a_3=b_{12}=d_{13}=d_{31}=d_{32}'=p_{13}'''=0 $ & $b_{11} d_{23}$ & T4(4)\\
(12) &  8  &  $a_1=a_2=a_3=b_{13}=b_{22}=b_{23}=i_{32}=o_{123}=0 $ & $b_{21} d_{21}$ & T2(3)\\
(13) &  8  &  $a_1=a_2=a_3=b_{11}=b_{32}=b_{33}=g_{32}=o_{123}=0 $ & $b_{22}$ & T2(3)\\
(14) &  8  &  $a_1=a_2=a_3=b_{21}=d_{12}=d_{23}=e_{31}=g_{13}=0 $ & $b_{33}$ & T2(3)\\
(15) &  8  &  $a_1=a_2=a_3=b_{32}=b_{33}=d_{21}'=g_{23}=o_{123}=0 $ & $b_{11} b_{22} d_{12}$ & T2(3)\\
(16) &  8  &  $a_1=a_2=a_3=b_{22}=b_{23}=d_{21}'=d_{31}'=o_{123}=0 $ & $b_{33}$ & T2(3)\\
(17) &  8  &  $a_1=a_2=a_3=b_{11}=b_{33}=d_{32}''=g_{23}=o_{123}=0 $ & $b_{13} b_{23} b_{22}$ & T2(3)\\
(18) &  8  &  $a_1=a_2=a_3=b_{22}=b_{33}=d_{31}''=g_{13}=o_{123}=0 $ & $b_{11} d_{21}$ & T2(3)\\
(19) &  9  &  $a_1=a_2=a_3=b_{11}=b_{12}=b_{33}=d_{32}''=g_{12}=g_{23}=0 $ &  & T2(3)\\
\hline
\multicolumn{5}{l}{}\\[-1ex]
\multicolumn{5}{@{}l}{$(^{*})\,\,d_{31}d_{32}(b_{11}b_{22}+b_{32}d_{31}'')\neq 0$}
\end{tabular}
\end{table}

\bigskip\newpage
\noindent{\bf Theorem 8.} {\it An LV3 is integrable with two first integrals
one of which being of the type of Theorem 2(3)(excluding those included
in Theorem 7) in the cases of Table 6.} 
{\small
\begin{longtable}{|c|c|l|l|l|}
  \caption{Integrability conditions of Theorem 8.}\\
  \hline  
   st. &  n & conditions =0&  conditions $\neq  0$ & f.i. type\\
  \hline
  \endfirsthead
  \caption[]{Integrability conditions of Theorem 8 (continued).}\\
  \hline  
   st. &  n & conditions =0&  conditions $\neq  0$ & f.i. type\\
  \hline
  \endhead
  \hline
  \endfoot
(1) &  5  &  $a_1=a_2=a_3=g_{12}=i_{23}'''=0$ & $d_{23}j_{13}$ & Q\\
(2) &  6  &  $a_1=c_{23}=b_{11}=b_{32}=e_{12}=g_{23}=0 $ & $b_{33}d_{23}d_{32}d_{13}j_{12}$ & Q\\
(3) &  6  &  $a_1=c_{23}=b_{11}=b_{21}=b_{31}=b_{32}=0$     &  (*) & P1(2),T2(3)\\
(4) &  6  &  $a_1=c_{23}=b_{11}=g_{13}=d_{12}'=g_{23}=0$ & $a_2a_3d_{23}d_{32}d_{13}d_{23}''$ & Q\\
(5) &  6  &  $a_1=c_{23}=b_{11}=b_{32}=o_{123}=g_{23}=0$ & $d_{23}d_{32}d_{13}$ & Q\\
(6) &  6  &  $a_1=c_{23}=b_{11}=b_{12}=d_{32}'=g_{23}=0$ & $d_{13}d_{23}''j_{12}$ & Q\\
(7)  &  6  &  $a_1=c_{23}=b_{11}=b_{13}=d_{23}'=g_{23}=0$ & $b_{12}d_{12}d_{32}''j_{13}$ & Q\\
(8)  &  6  &  $a_1=c_{23}=b_{11}=g_{23}=n_{23}=o_{123}=0$ & $b_{22}b_{33}d_{23}d_{32}d_{12}d_{32}''$ & Q\\
(9)  &  6  &  $a_1=c_{23}=b_{11}=b_{12}=b_{32}=g_{23}=0$ & $b_{22}b_{33}b_{13}d_{13}$ & T2(7)\\
(10) &  6  &  $a_2=c_{13}=b_{22}=g_{23}=g_{13}=r_{31}=0 $ & $b_{33}$ & Q\\
(11) &  6  &  $a_1=c_{23}=b_{11}=n_{23}'=r_{23}=g_{23}=0 $ & $b_{22} b_{33} d_{13} d_{32} h_{23}'$ & Q\\
(12) &  6  &  $a_1=c_{23}=b_{11}=b_{13}=b_{23}=g_{23}=0 $ & $b_{12}d_{32}$ & P1(8)\\
(13) &  6  &  $a_1=c_{23}=b_{11}=g_{23}=g_{13}=r_{23}=0 $ & $a_2 a_3d_{13} d_{12}$ & Q\\
(14) &  6  &  $a_1=a_2=a_3=g_{12}=g_{23}=g_{31}=0$ & $b_{21}d_{23}d_{32}$ & Q\\
(15) &  6  &  $a_1=a_2=a_3=b_{13}=b_{23}=g_{23}=0 $ & (**) & T2(3)\\
(16) &  7  &  $a_1=c_{23}=b_{11}=b_{32}=b_{12}=g_{12}=g_{23}=0$ & $b_{33}b_{21}d_{23}d_{32}$ & Q\\
(17) &  7  &  $a_1=c_{12}'''=c_{13}''=h_{31}'=p_{32}=g_{32}=g_{31}=0 $ & $b_{11} b_{22} b_{33} d_{13} d_{12}$ & T2(3)\\
(18) &  7  &  $a_1=a_2=a_3=d_{32}'=d_{12}'=n_{31}=q_{31}'''$ & (***) & Q\\
(19) &  7  &  $a_1=a_2=a_3=g_{23}=n_{23}=n_{23}'=p_{13}^{iv}=0$ & $b_{12}b_{13}b_{22}d_{23}$ & T4(1)\\
(20) &  7  &  $a_1=a_2=a_3=b_{12}=b_{13}=n_{23}=g_{32}=0 $ & $b_{22} b_{33} d_{32} d_{32}''$ & T4(1)\\
(21) &  7  &  $a_1=a_2=a_3=b_{12}=b_{32}=b_{33}=g_{23}=0 $ & $b_{21} b_{13}$ & T2(3)\\
(22) &  7  &  $a_1=a_2=a_3=d_{21}'=d_{31}'=n_{23}=q_{23}'''=0 $ & $b_{33} d_{32} d_{32}'' i_{31}''$  & Q\\
(23) &  7  &  $a_1=a_2=a_3=b_{13}=b_{22}=d_{23}'=g_{23}=0 $ & (****) & Q\\
(24) &  7  &  $a_1=a_2=a_3=b_{33}=g_{31}=g_{32}=q_{21}'=0 $ & $b_{13} b_{23} d_{12} g_{12}''$ & Q\\
(25) &  8  &  $c_{23}=c_{31}'=b_{11}=b_{21}=b_{31}=b_{32}=d_{12}'=d_{13}=0$ & $a_1a_2a_3b_{33}d_{23}\neq 0$ & P1(2),T2(3)\\
(26) &  8  &  $c_{12}=c_{23}=b_{11}=r_{21}=r_{31}=r_{23}'=r_{32}'=n_{13}'=0$ & $b_{22}b_{33}$ & T4(37)\\
(27) &  8  &  $a_1=c_{23}=b_{11}=b_{13}=b_{21}=b_{31}=d_{23}'=j_{13}=0$ & $a_2a_3$ & Q\\
(28) &  8  &  $a_1=a_2=a_3=b_{33}=e_{31}=e_{32}=g_{13}=g_{23}=0$ & $b_{23}b_{31}b_{21}d_{23}$ & T2(3)\\
(29) &  8  &  $a_1=a_2=a_3=b_{13}=b_{22}=b_{23}=d_{31}''=g_{13}=0 $ & $b_{21}$ & T2(3)\\
(30) &  8  &  $a_1=a_2=a_3=b_{22}=d_{21}''=d_{13}'=g_{12}=g_{23}=0 $ & $g_{31}'$ & T2(3)\\
(31) &  8  &  $a_1=a_2=a_3=b_{11}=b_{22}=b_{33}=g_{23}=g_{13}=0 $ & $g_{12}''$ & T2(3)\\
(32) &  8  &  $a_1=a_2=a_3=b_{11}=b_{22}=b_{33}=g_{23}=n_{23}'=0 $ & $b_{23} b_{21} g_{12}''$ & Q\\
(33) &  9  &  $c_{12}''=c_{23}''=b_{12}=b_{22}=b_{32}=d_{23}=e_{31}=g_{23}=r_{31}=0 $ & $b_{33}$ & Q\\
(34) &  9  &  $a_1=c_{23}=c_{31}'''=b_{11}=b_{31}=d_{12}''=d_{32}=d_{23}=d_{13}=0 $ & $b_{21} b_{33}$ & Q\\
(35) &  9  &  $a_1=a_2=a_3=b_{33}=d_{12}'=d_{21}''=d_{32}''=g_{32}=g_{21}'''=0 $ & $b_{11}$ & Q\\
(36) &  9  &  $a_1=a_2=a_3=d_{12}=d_{13}=d_{21}=d_{31}=n_{32}=j_{12}'=0 $ & $b_{33}d_{13}$ & Q\\
(37) &  9  &  $a_1=a_2=a_3=b_{22}=b_{23}=d_{21}''=d_{13}'=g_{23}=g_{31}'''=0 $ & $b_{12}$ & Q\\
(38) &  9  &  $a_1=a_2=a_3=b_{22}=b_{33}=d_{31}''=g_{23}=g_{21}'''=g_{31}'''=0 $ & $b_{12} b_{23}$ & Q\\
\end{longtable}
\vspace{-3ex}
\noindent (*) $b_{22}b_{33}d_{23}(b_{22}b_{13}+b_{33}b_{12}-b_{23}b_{12})\neq 0\quad$
\noindent (**) $d_{12}d_{21}d_{32}(b_{12}b_{21}-b_{11}b_{22})\neq 0$\\
\noindent (***) $b_{11}d_{13}d_{13}''i_{12}''(2d_{13}-b_{33})\neq 0\qquad\qquad$
\noindent (****) $b_{12} g_{31}'' (b_{12} b_{21}+b_{11} b_{32})\neq 0$.
}

\bigskip\newpage
\noindent{\bf Theorem 9.} {\it An LV3 is integrable with two first integrals
one of which being of the type of Theorem 2(4) in the cases of Table 7.}
{\small
\begin{longtable}{|c|c|l|l|l|}
  \caption{Integrability conditions of Theorem 9.}\\
  \hline  
   st. &  n & conditions =0&  conditions $\neq  0$ & f.i. type\\
  \hline
  \endfirsthead
  \caption[]{Integrability conditions of Theorem 9 (continued).}\\
  \hline  
   st. &  n & conditions =0&  conditions $\neq  0$ & f.i. type\\
  \hline
  \endhead
  \hline
  \endfoot
(1) &  6  &  $b_{11}=b_{12}=b_{21}=b_{31}=b_{32}=r_{23}=0 $ & $a_2 b_{22} b_{33}$ & T2(7)\\
(2) &  6  &  $b_{13}=b_{21}=b_{23}=b_{31}=r_{12}=r_{23}=0 $ & $a_1a_2a_3$ & T2(4)\\
(3) &  6  &  $b_{13}=b_{21}=b_{23}=e_{31}=r_{12}=r_{32}=0 $ & $a_2 b_{11}$ & Q\\
(4) &  6  &  $c_{12}=c_{32}'=b_{13}=b_{23}=o_{123}=r_{12}=0 $ & $a_1a_2 a_3b_{11} d_{21}$ & Q\\
(5) &  6  &  $a_1=c_{23}=b_{11}=g_{23}=g_{13}=r_{23}=0$ & $b_{22}d_{12}$ & P1(3)\\
(6) &  7  &  $c_{23}'''=b_{12}=b_{31}=b_{32}=e_{21}=e_{23}=r_{31}=0 $ & $a_2 a_3 c_{12}''$ & Q\\
(7) &  7  &  $c_{31}''=b_{12}=b_{13}=b_{31}=b_{32}=e_{21}=r_{23}=0 $ & $a_1 a_2 a_3c_{12}''$ & Q\\
(8) &  7  &  $c_{23}''=b_{23}=b_{31}=b_{32}=b_{12}=e_{21}=r_{13}=0 $ & $a_1 a_2 a_3c_{12}''$ & Q\\
(9) &  8  &  $c_{12}''=b_{11}=b_{21}=b_{23}=b_{31}=d_{12}=d_{13}=r_{32}=0 $ & $a_1a_2$ & P1(2)\\
(10) &  8  &  $c_{31}''=b_{11}=b_{21}=b_{31}=b_{32}=d_{12}=d_{13}=r_{23}=0 $ & $a_1a_3 b_{33}$ & P1(2)\\
(11) &  8  &  $c_{13}''=b_{11}=b_{13}=b_{21}=b_{23}=b_{31}=r_{23}=r_{12}''=0 $ & $a_1a_2a_3$ & P1(2)\\
(12) &  8  &  $c_{23}=b_{12}=b_{22}=b_{31}=b_{32}=d_{23}=e_{21}=s_{13}''=0 $ & $b_{23}$ & P1(6)\\
(13) &  8  &  $c_{12}''=c_{13}=b_{12}=b_{21}=b_{23}=b_{32}=e_{31}=r_{13}'=0 $ & $a_1a_2a_3$ & T5(19)\\
(14) &  8 & $c_{12}'''=c_{32}=b_{12}=b_{13}=b_{32}=d_{21}'=d_{23}''=d_{31}'=0 $ & $a_1a_2a_3$ & T5(14)\\
(15) &  8  &  $c_{23}=c_{13}'''=b_{12}=b_{13}=b_{23}=d_{21}'=d_{32}''=g_{32}=0 $ & $a_1a_2a_3$ & Q\\
(16) &  8  &  $c_{12}=c_{23}'''=b_{12}=b_{13}=b_{21}=b_{23}=b_{32}=d_{31}'=0 $ & $a_1a_2a_3$ & T5(43)\\
(17) &  8  &  $a_1=b_{11}=b_{21}=b_{32}=b_{31}=d_{13}=d_{12}'=r_{23}=0 $ & $|a_3|+|b_{33}|$ & P1(2)\\
(18) &  8  &  $a_3=b_{13}=b_{21}=b_{23}=b_{33}=d_{32}=e_{31}=r_{12}=0 $ & $a_2$ & P1(2)\\
(19) &  8 & $a_3=b_{33}=b_{13}=b_{23}=b_{21}=d_{31}=d_{32}''=r_{12}=0 $ & $a_1 a_2$ & P1(2)\\
(20) &  8  &  $a_1=c_{23}=b_{11}=b_{21}=b_{31}=d_{13}=j_{13}=r_{32}=0 $ & $b_{22}$ & P1(3)\\
(21) &  8  &  $a_2=b_{12}=b_{21}=b_{22}=b_{31}=b_{32}=d_{23}=r_{31}=0 $ & $a_1a_3$ & P1(2)\\
(22) &  8  &  $a_1=b_{11}=b_{21}=b_{31}=b_{32}=d_{13}=d_{12}''=s_{23}=0 $ & $a_2a_3$ & P1(2)\\
(23) &  9  &  $a_3=c_{12}'''=b_{12}=b_{13}=b_{23}=b_{31}=b_{33}=d_{21}'=d_{32}=0 $ & $|a_1|+|b_{11}|+|b_{22}|$ & Q\\
(24) &  9  &  $c_{23}=c_{31}'=b_{12}=b_{13}=b_{23}=b_{32}=b_{33}=d_{31}=r_{12}=0 $ & $b_{11} b_{22}$ & T2(5)\\
(25) &  9  &  $c_{12}''=c_{13}=b_{11}=b_{21}=b_{23}=b_{31}=b_{32}=d_{12}=d_{13}=0 $ & $a_1a_2a_3$ & P1(2)\\
(26) &  9  &  $c_{12}=c_{23}=b_{11}=b_{21}=b_{31}=b_{32}=d_{12}=d_{13}=d_{23}''=0 $ & $a_1a_2a_3b_{22}b_{33}$ & P1(2,3)\\
(27) &  9  &  $c_{13}=c_{23}=b_{11}=b_{21}=b_{23}=b_{31}=d_{12}=d_{13}''=d_{32}''=0 $ & (*) & P1(2,3)\\
(28) &  9  &  $c_{23}=c_{21}'=b_{12}=b_{22}=b_{31}=b_{32}=d_{21}=d_{23}=j_{12}''=0 $ & $|a_1|+|b_{11}|+|b_{33}|$ & T4(4)\\
(29) &  9  &  $c_{12}=c_{13}=b_{12}=b_{21}=b_{22}=b_{32}=d_{23}=d_{31}'=r_{13}=0 $ & $b_{11}$ & Q\\
(30) &  9  &  $c_{21}'=c_{23}=b_{12}=b_{31}=b_{22}=b_{32}=d_{23}=d_{21}''=r_{31}=0 $ & $b_{13}$ & Q\\
(31) &  9  &  $a_3=c_{21}''=b_{23}=b_{21}=b_{33}=b_{12}=b_{13}=d_{32}=d_{31}''=0 $ & $|a_1|+|b_{11}|$ & P1(2)\\
(32) &  9  &  $a_1=c_{23}''=b_{11}=b_{13}=b_{21}=b_{23}=b_{31}=b_{32}=d_{12}''=0 $ & $|a_2|+|b_{33}|$ & P1(2)\\
(33) &  9  &  $a_1=c_{23}=b_{11}=b_{21}=b_{31}=d_{13}=d_{12}'=d_{23}'=n_{32}'=0 $ & $b_{22}b_{33}$ & P1(3)\\
(34) &  9  &  $a_1=c_{23}''=b_{11}=b_{21}=b_{23}=b_{31}=b_{32}=d_{12}=d_{13}''=0 $ & $|a_2|+|b_{33}|$ & P1(2)\\
(35) &  9  &  $a_3=c_{12}''=b_{12}=b_{13}=b_{21}=b_{23}=b_{33}=d_{32}=e_{31}=0 $ & $|a_2|+|b_{22}|$ & P1(2)\\
(36) &  9  &  $a_2=c_{13}'''=b_{12}=b_{22}=b_{31}=b_{32}=d_{21}=d_{23}''=e_{13}=0 $ & $|a_3|+|b_{33}|$ & P1(2)\\
(37) &  9  &  $a_1=c_{23}''=b_{11}=b_{21}=b_{23}=b_{31}=b_{32}=d_{12}=d_{13}''=0 $ & $|a_3|+|b_{33}|$ & P1(2)\\
(38) &  9  &  $a_2=c_{13}=b_{12}=b_{22}=b_{32}=d_{23}=e_{31}=g_{23}=r_{13}=0 $ & $|a_1|+|b_{11}|+|b_{33}|$ & Q\\
(39) &  9  &  $a_2=c_{13}'''=b_{12}=b_{13}=b_{21}=b_{22}=b_{32}=d_{31}'=d_{23}=0 $ & $|a_1|+|b_{11}|+|b_{33}|$ & Q\\
(40) &  9  &  $a_3=c_{21}'''=b_{12}=b_{13}=b_{23}=b_{32}=b_{33}=e_{21}=e_{31}=0 $ & $|a_1|+|b_{11}|+|b_{22}|$ & Q\\
(41) &  9  &  $a_3=c_{12}'''=b_{12}=b_{13}=b_{23}=b_{31}=b_{33}=d_{21}'=d_{32}''=0 $ & $a_1a_2$ & Q\\
\end{longtable}
\vspace{-3ex}
\noindent (*) $(a_2+b_{22})(b_{22}+b_{33})\neq 0$
}

\bigskip\newpage
\noindent{\bf Theorem 10.} {\it The LV3 has two first integrals with at least one being formed 
by one linear algebraic solution not included in Theorems 6-9 in the cases of Table 8.}
\begin{table}[h!]
\vspace{-1em}
\caption{Integrability conditions of Theorem 10.}
\vspace{1em}
\small
\centering
\begin{tabular}{|c|c|l|l|l|}
\hline  
 st. &  n & conditions =0&  cond. $\neq  0$ & f.i. type\\
\hline
(1)  &  5  &  $a_{1}=b_{11}=b_{12}=b_{32}=d_{13}''$ & $a_3b_{31}b_{33}$ & T2(7),Q\\
(2)  &  6  &  $b_{11}=b_{12}=b_{21}=b_{31}=b_{32}=r_{23}'=0$ & $a_2a_3c_{23}$ & T2(7),T5(1)\\
(3)  &  6  &  $a_{1}=c_{32}'=b_{11}=b_{13}=b_{23}=g_{32}'=0$  & $b_{12}b_{22}d_{12}d_{32}''$ & 2T2(7)\\
(4)  &  7  &  $c_{23}''=b_{23}=b_{31}=b_{32}=b_{13}=d_{21}=d_{12}=0$  & $a_1a_2a_3c_{12}$ & T2(5),Q\\
(5)  &  8  &  $c_{23}=b_{12}=b_{13}=b_{22}=b_{31}=b_{32}=e_{21}=d_{23}''=0$  & $c_{12}$ & 2P1(2)\\
(6) &  8  &  $c_{31}'=b_{12}=b_{13}=b_{22}=b_{23}=b_{32}=b_{33}=d_{31}=0$ & $l_{32}$ & T2(5),Q\\
(7) &  8  &  $c_{21}'=c_{23}=b_{12}=b_{13}=b_{22}=b_{32}=d_{23}=d_{21}=0$  & $b_{21}b_{23}$ & T2(11),Q\\
(8) &  8  &  $c_{12}=c_{23}=b_{12}=b_{22}=b_{31}=b_{32}=d_{21}=d_{23}=0$  & $a_1a_2a_3b_{11}$ & T2(11),Q\\
(9) &  8  &  $c_{32}=c_{13}''=b_{23}=b_{31}=b_{32}=b_{13}=d_{21}=d_{12}=0$  & $b_{11}$ & T2(5),T5(3)\\
(10) &  8  &  $a_{1}=b_{31}=b_{32}=b_{11}=b_{12}=b_{23}=b_{21}=d_{13}=0$  & $a_2a_{3}$ & T2(7),Q\\
(11) &  8  &  $a_{1}=c_{23}=b_{11}=b_{13}=b_{21}=b_{31}=d_{12}=d_{23}'=0$  & $b_{22}$ & P1(3),Q\\
(12) &  8  &  $a_{1}=c_{32}'''=b_{11}=b_{12}=b_{32}=d_{13}'=g_{12}=g_{32}'''=0$  & $a_2a_{3}$ & T2(6),Q\\
(13) &  8  &  $a_{1}=c_{32}'''=b_{11}=b_{12}=b_{23}=b_{32}=e_{13}=g_{32}'''=0$  & $a_2a_{3}$ & T2(6),Q\\
(14) &  8  &  $a_{1}=c_{32}'''=b_{11}=b_{12}=b_{23}=b_{32}=d_{13}'=g_{32}'''=0$  & $a_2a_{3}$ & T2(6),Q\\
(15) &  9  &  $c_{12}=c_{13}=b_{11}=b_{21}=b_{23}=b_{31}=d_{12}=d_{13}''=n_{23}=0$  & $a_1a_{2}a_3$ & P1(2),Q\\
(16) &  9  &  $a_{1}=c_{32}'''=b_{11}=b_{13}=b_{21}=b_{23}=b_{31}=b_{32}=d_{12}''=0$  & $|a_3|+|b_{33}|$ & P1(2),Q\\
(17) &  9  &  $a_{1}=c_{23}''=b_{11}=b_{21}=b_{31}=d_{12}''=d_{13}''=d_{23}=d_{32}=0$  & $a_2a_3$ & T2(5),Q\\
\hline
\end{tabular}
\end{table}

\noindent{\bf Theorem 11.} {\it The LV3 has two first integrals with quadratic 
algebraic solutions in the cases of Table 9.}
\begin{table}[h!]
\vspace{-1em}
\caption{Integrability conditions of Theorem 11.}
\vspace{1em}
\small\centering
\begin{tabular}{|c|c|l|l|l|}
\hline  
 st. &  n & conditions =0&  cond. $\neq  0$ & f.i. type\\
\hline
(1) &  7  &  $c_{23}=b_{12}=b_{13}=d_{23}'=d_{32}'=g_{23}=r_{13}=0$ & (*) & Q,Q\\
(2) &  7  &  $c_{23}'''=b_{31}=b_{32}=d_{12}=d_{21}=e_{23}=n_{13}'=0$ & $a_1a_2a_3b_{11}b_{33}$ & Q,Q\\
(3) &  7  &  $c_{23}=c_{21}'=d_{21}=d_{12}=d_{31}=r_{23}=r_{13}=0$  &  $b_{11}b_{22}b_{33}d_{23}$ & T4(12),Q\\
(4) &  7  &  $c_{23}=c_{21}'=b_{12}=g_{32}=n_{12}=r_{23}=n_{31}=0$ & $d_{32}d_{31}''$ & T4(12),T4(28)\\
(5) &  7  &  $c_{12}''=c_{13}'=b_{23}=b_{31}=e_{12}=e_{21}=e_{32}=0$ & $b_{13}b_{33}d_{13}''$ & T5(30),Q\\
(6) &  8  &  $c_{12}=c_{23}=b_{13}=r_{21}=r_{31}=r_{23}'=r_{32}'=n_{13}'=0$ & $b_{22}b_{33}$ & T4(37),T4(1)\\
(7) &  8  &  $c_{12}''=c_{13}''=b_{31}=b_{21}=r_{12}=r_{13}=r_{23}'=r_{32}'=0$ & $a_1a_2a_3b_{11}b_{12}$ & T4(2),T5(46)\\
(8) &  8  &  $c_{12}=c_{13}=b_{21}=b_{23}=b_{31}=g_{13}=n_{12}=n_{23}'=0$ & $b_{22}b_{33}$ & T4(1), T4(2)\\
(9) &  8  &  $a_3=c_{12}'''=b_{12}=b_{21}=b_{32}=b_{33}=e_{31}=g_{12}'''=0$ & $a_1a_2$ & Q,Q\\
(10) &  8  &  $a_2=c_{31}'''=b_{21}=b_{22}=b_{31}=d_{13}'=d_{23}''=n_{23}'=0$ & $a_1a_3$ & T5(4),Q\\
(11) &  8  &  $a_{3}=c_{12}''=b_{32}=b_{33}=d_{21}=d_{12}=g_{32}'=g_{12}''=0$  &  $a_1$ & T5(56),T5(57)\\
(12) &  9  &  $c_{12}''=c_{13}'=b_{11}=b_{21}=b_{23}=b_{31}=d_{12}=d_{13}''=r_{32}'=0$ & $a_1a_2a_3$ & Q,Q\\
(13) &  9  &  $c_{21}'=c_{23}'''=b_{11}=b_{21}=b_{23}=b_{31}=d_{13}=d_{12}'=n_{23}=0$ & $a_1a_2a_3$ & T5(5), Q\\
(14) &  9  &  $c_{21}'=c_{31}'=b_{12}=b_{13}=b_{22}=b_{32}=d_{21}''=d_{23}''=n_{31}=0$  & $a_1a_2a_3b_{33}$ &T4(4),T4(27)\\ 
(15) &  9  &  $c_{13}=c_{23}=b_{11}=b_{21}=b_{23}=b_{31}=d_{12}=d_{13}''=n_{23}=0$ & $b_{22}b_{33}$ & Q,Q\\
(16) &  9  &  $c_{12}''=c_{23}=b_{11}=b_{21}=b_{31}=d_{12}=d_{13}=r_{23}'=n_{23}=0$  &  $a_1a_2a_3$ & T5(10),T5(46)\\
\hline
\multicolumn{5}{l}{} \\[-1ex]
\multicolumn{5}{@{}l}{(*) $a_1b_{11}d_{21}d_{31}d_{23}''d_{32}''\neq 0$.}
\end{tabular}
\end{table}

We note, before concluding, that Theorem 4(6) and Theorem 8(1) with the additional 
conditions $b_{11}=b_{22}=b_{33}=0$ and Theorem 8(31--32) are Theorem 6(5--6) and Theorem 7(4) and 7(7) 
respectively of [4] in the case of ABC systems. Moreover, we note that to the above 
mentioned cases of Theorem 11, one must add the case obtained by Moulin--Ollagnier [30] 
for the ABC system which is formed with the first integral of Theorem 4(5) and an other 
one which is not of Darboux type (see Ref. 31 for more details).  

\section{Conclusion}

The use of Darboux method for the LV3 has been rather efficient to find first 
integrals and to detect integrable cases. The main reason of this efficiency
must be found in the fact that the problem of searching a first integral has
been reduced here to the search of only one algebraic solution. (Theorem 3 is 
when a quadratic algebraic solution can be factorised in two linear ones). In total we have 
found 366 cases, among which 172 are cases of a single first integral and 194 when
two first integrals coexist. As predicted, we see that for the existence of quadratic
invariant solutions  and the corresponding first integrals, we require a larger number 
of conditions than for linear algebraic solutions. In
fact for having a single first integral formed by planes we can need only a maximum of 3 
conditions in total (Theorem 2 (1)-(4)), whereas it is required 4 to 8 conditions 
when at least there is one second degree solution. The cases of integrability  require at 
least 4 conditions when both first integrals are linear (in fact one case with 4 conditions 
and 5 with 5 conditions). Otherwise their number can grow now to 9 conditions. 
In general the number of conditions required is the number corresponding 
to the more constrained first integral increased by one. The number of essential
parameters of the LV3 system being 9 (8 if we normalize the time),  we omitted to consider 
the cases having more than 9 conditions. The present work can be completed with the 
cases where the planes $x=0$, $y=0$, $z=0$ and $f_4=0$ are replaced by other planes and polynomials of 
greater degree, by exponential factors (see Cair\'o and Llibre [4]) and with time-dependent first integrals.

\label{lastpage}


\begin{thebibliography}{99}
\small

\bibitem{[1]}  Lotka A.J.,
Analytical Note on Certain Rhythmic Relations in Organic Systems,
{\it Proc. Natl. Acad. Sci. U.S.}, 1920, V.6, 410--415.

\bibitem{[2]}  Volterra V., Le\c{c}ons sur la Th\'{e}orie 
Math\'{e}matique de la Lutte pour la vie, Gauthier 
Villars, Paris, 1931.

\bibitem{[3]}  May R.M. and Leonard W.J., Nonlinear Aspects
of Competition Between Three Species, {\it Siam J. Appl. Math.}, 
1975, V.29, 243--253.

\bibitem{[4]}  Cair\'{o} L. and Llibre J., Darboux Integrability for the 
3-Dimensional Lotka--Volterra Systems, {\it J. Physics A, Math. and Gen.}, 
2000, V.33, 2395--2406. 

\bibitem{[5]}  Lamb W.E., Theory of an Optical Maser,
{\it Phys Rev. A}, 1964, V.134, 1429. 

\bibitem{[6]}  Laval G. and Pellat R., Plasma Physics,
in Proceedings of Summer School of Theoretical Physics, Gordon and 
Breach, New York, 1975.

\bibitem{[7]}  Busse F.H., Transitions to Turbulence Via 
the Statistical Limit Cycle Route, in Synergetics, Springer-Verlag,
 Berlin, 1978, p.\ 39.

\bibitem{[8]}  Lupini R., Spiga G., Chaotic Dynamics of 
Spatially Homogeneous Gas Mixtures, {\it Phys. Fluids.}, 1988, V.31,
2048--2051. 

\bibitem{[9]} Bogoyavlensky O.I., Integrable 
Discretizations of the KdV Equations, {\it Phys. Lett. A}, 1988, V.134,
34.

\bibitem{[{10}]}  Noonburg V.W., A Neural Network Modeled
 by an Adaptive Lotka-Volterra System, {\it SIAM J. Appl. Math.},
1989, V.49, 1779. 

\bibitem{[{11}]}  Brenig L., Complete Factorisation and Analytic
 Aolutions of Generalized Lotka-Volterra Equations, {\it Phys. 
Lett.}, 1988, V.133, 378--382.

\bibitem{[{12}]}  Brenig L. and Goriely A., Universal Canonical
 Forms for Time-Continuous Dynamical Systems, {\it Phys. Rev. A}, 
1989, V.40, 4119--4122.

\bibitem{[{13}]} Ku\`s M., Integrals of Motion for the Lorenz
 System, {\it J.~Phys.~A: Math.~Gen.}, 1983, V.16, L689--L691.
 
\bibitem{[{14}]}  Cair\'{o} L. and Feix M.R., Families of
 Invariants of the Motion for the Lotka-Volterra Equations: the
 Linear Polynomials Family, {\it J.~Math.~Phys.}, 1992a, V.33,
 2440--2455.

\bibitem{[{15}]}  Cair\'{o} L., Feix M.R., Hua D.D. and Bouquet S.,
Hamiltonian Method and Invariant Search for 2D Quadratic
Systems, {\it J.~Phys.~A: Math.~Gen.}, 1993a, V.26, 4371--4386. 

\bibitem{[{16}]}  Hua D.D., Cair\'{o} L. and Feix M.R.,
Time-Independent Invariants for the Quadratic System, 
{\it J.~Phys.~A: Math.~Gen.}, 1993b, V.26, 7097--7114.

\bibitem{[{17}]}  Bountis T., Ramani A., Grammaticos B. and 
Dorizzi B., `On the Complete and Partial Integrability of 
Non-Hamiltonian Systems, {\it Physica A}, 1984, V.128, 268--288. 
 
\bibitem{[{18}]}  Goriely A., Investigation of Painlev\'e
 Property Under Time Singularities Transformations, 
{\it J. Math. Phys.}, 1992, V.33, 2728--2742.

\bibitem{[{19}]}  Steeb W.H., Continuous Symmetries of the 
Lorenz Model and the Rikitake Two-Disc Dynamo System, {\it J.
 Phys. A: Math. Gen.}, 1982, V.15, L389--L390.

\bibitem{[{20}]}   Almeida M.A., Magalh\~aes M.E. and Moreira I.C.,
Lie Symmetries and Invariants of the Lotka--Volterra System, 
{\it J. Math. Phys.}, 1995, V.36, 1854--1867.

\bibitem{[{21}]}  Strelcyn J.M. and Wojciechowski S., 
A Method of Finding Integrals for Three-Dimensional Dynamical
 Systems, {\it Phys. Lett. A}, 1988, V.133, 207--212.

\bibitem{[{22}]}  Grammaticos B., Moulin-Ollagnier J.,
 Ramani A., Strelcyn J.M. and Wojciechowski S., 
Integrals of Quadratic Ordinary Differential Equations in $R^3$: the
 Lotka-Volterra System, {\it Physica A}, 1990, V.163, 683--722.

\bibitem{[{23}]} Cair\'{o} L. and Feix M.R., On the Hamiltonian 
Structure of the 2D ODE Possessing an Invariant, {\it J.~Phys.~A:
 Math.~Gen.}, 1992b, V.25, L1287--L1293.

\bibitem{[{24}]}  Hua D.D., Cair\'{o} L. and Feix M.R., 
`A General Time-Fependent Invariant for and Integrability of
 the Quadratic System, {\it Proc. of the Royal Soc. London A}
 1993c, V.443, 643--650.

\bibitem{[{25}]}  Darboux G., M\'emoire sur les \'equations 
diff\'erentielles alg\'ebriques du premier ordre et du premier degr\'e 
{\it (M\'elanges), Bull. Sci. Math.}, 1878,  {\bf 60--96}; 123--144; 151--200.

\bibitem{[{26}]}  Schlomiuk D., Algebraic Particular Integrals, 
Integrability and the Problem of the Center, {\it Trans. Amer. Math.
Soc.} 1993, V.338, 799--841.
  
\bibitem{[{27}]}  Cair\'o L., Llibre J. and  Feix M.R., Darboux Method
and Search of Invariants for the Lotka-Volerra and Complex Quadratic Systems,
{\it J. of Math. Phys.}, 1999, V.40, 2074--2091.

\bibitem{[{28}]}  Cair\'o L., Llibre J. and  Feix M.R., Integrability and
 Algebraic Aolutions for Planar Polynomial Differential Systems with Emphasis
 on the Quadratic Systems, {\it Resenhas IME--USP}, 2000, V.4, 127--161.

\bibitem{[{29}]} Cair\'o L. and Llibre J.,  Integrability and Algebraic
Solutions for the 2-d Lotka--Volterra System, 
in Dynamical systems, Plasmas and Gravitation. 
Springer--Verlag Monograph in Physics, 1999, V.518, 243--254.

\bibitem{[{30}]} Moulin-Ollagnier J., 
Rational Integration of the Lotka--Volterra
 System, {\it Bull. Sci. Math.}, 1999, V.123, 437--466. 

\bibitem{[{31}]} Cair\'o L. and Llibre J., Algebraic Integrability
for a Class of 3-Dimensional Lotka--Volterra Systems, 
preprint Univ.\ Orleans, 2000.

\end{thebibliography}
\end{document}